\documentclass[preprint,pre,aps,amsfonts,amsmath,showpacs,superscriptaddress,nofootinbib]{revtex4}
\pdfoutput=1
\usepackage{amsmath,amssymb,amscd}
\usepackage{graphicx}
\usepackage{bm}

\newcommand{\e}{{\rm e}}

\begin{document}
\title{On the Gap and Time Interval between the First Two Maxima of Long Continuous Time Random Walks}
\author{Philippe Mounaix}
\email{philippe.mounaix@cpht.polytechnique.fr}
\affiliation{Centre de Physique Th\'eorique, UMR 7644 du CNRS, Ecole
Polytechnique, 91128 Palaiseau Cedex, France.}
\author{Gr\'egory Schehr}
\email{gregory.schehr@lptms.u-psud.fr}
\affiliation{Laboratoire de Physique Th\'eorique et Mod\`eles Statistiques, Universit\'e Paris-Sud, B\^at 100, 91405 Orsay Cedex, France.}
\author{Satya N. Majumdar}
\email{majumdar@lptms.u-psud.fr}
\affiliation{Laboratoire de Physique Th\'eorique et Mod\`eles Statistiques, Universit\'e Paris-Sud, B\^at 100, 91405 Orsay Cedex, France.}
\date{\today}
\begin{abstract}
We consider a one-dimensional continuous time random walk (CTRW) on a fixed time interval $T$ where at each time step the walker waits a random time $\tau$, before performing a jump drawn from a symmetric continuous probability distribution function (PDF) $f(\eta)$, of L\'evy index $0 < \mu \leq 2$. Our study includes the case where the waiting time PDF $\Psi(\tau)$ has a power law tail, $\Psi(\tau) \propto \tau^{-1 - \gamma}$, with $0< \gamma < 1$, such that the average time between two consecutive jumps is infinite. The random motion is sub-diffusive if $\gamma < \mu/2$ (and super-diffusive if $\gamma > \mu/2$). We investigate the joint PDF of the gap $g$ between the first two highest positions of the CTRW and the time $t$ separating these two maxima. We show that this PDF reaches a stationary limiting joint distribution $p(g,t)$ in the limit of long CTRW, $T \to \infty$. Our exact analytical results show a very rich behavior of this joint PDF in the $(\gamma, \mu)$ plane, which we study in great detail. Our main results are verified by numerical simulations. This work provides a non trivial extension to CTRWs of the recent study in the discrete time setting by Majumdar {\it et al.} ({\it J. Stat. Mech.} P09013, 2014).
\end{abstract}
\maketitle
%
%%%%%%%%%%%%%%%%%%%%
%
\section{Introduction and summary of main results}\label{sec1}

Recently, there has been a resurgence of interest in extreme value questions
for random walks and its variants, particularly in the physics literature \cite{AS2005,DB09,DB11,MMS2013,MMS2014,SLD2010}. This is motivated to a large extent by the observation that extreme value statistics (EVS) plays a crucial role in the physics of complex and disordered systems \cite{BM97, DM01}. Although the EVS of independent and identically distributed (i.i.d.) random
variables is very well understood, much less is known for strongly correlated variables, which is the case frequently encountered
in physics. From that point of view, random walks offer a very useful framework where the effects of correlations on EVS can be characterized
analytically. 

The statistics of the maximal displacement of a random walk (RW) after a given number of steps $n$ is now quite well understood, thanks for instance to the Pollaczeck-Spitzer formula~\cite{Poll,Spitz}, from which precise asymptotic estimates in the large $n$ limit can be obtained\ \cite{AS2005}. The full order statistics, i.e. the statistics of the $k^{\rm th}$ maximum for any integer $k$, was investigated in\ \cite{SM2012} in the case of ordinary RWs with a narrow jump distribution (characterized by a L\'evy exponent $\mu = 2$). Of particular interest is the {\it gap} between two consecutive maxima, which makes it possible to characterize the ``crowding'' of the walker positions near their maximum. Along this line we recently obtained a complete description of the joint distribution of the gap and time interval between the first two maxima of RWs not only for narrow jump distributions ($\mu =2$), but also for L\'evy flights characterized by a L\'evy exponent $0<\mu< 2$\ \cite{MMS2013,MMS2014}. Since the characteristic displacement of the walker after $n$ time steps grows like $n^{1/\mu}$, the results of\ \cite{MMS2013,MMS2014} are of interest for diffusive ($\mu =2$) and super-diffusive ($0<\mu <2$) processes only, leaving out the important case of {\it sub-diffusive} processes. Extending the analysis of\ \cite{MMS2013,MMS2014} to sub-diffusive RWs has been one of the motivations for this work.

A common model to describe sub-diffusive processes is the so called continuous time random walk (CTRW). In this model, the walker performs a usual random walk where two consecutive jumps are separated by a certain trapping time $\tau$. The trapping times are i.i.d. random variables drawn from a common distribution $\Psi(\tau)$ which has a power law tail $\Psi(\tau) \propto \tau^{-1-\gamma}$ with $0 < \gamma <1$. This model, initially suggested by Scher and Montroll in the setting of non-Gaussian transport in disordered electronic systems \cite{SM1975}, has been widely used to describe anomalous sub-diffusive dynamics in various complex systems phenomenologically\ \cite{BG90,MK2000}. Indeed, for $0<\gamma < 1$ the average trapping time between two successive jumps is infinite and, consequently, CTRWs are characterized by an anomalous growth of displacement $\propto t^{\gamma/\mu}$ which is indeed sub-diffusive if $\gamma < \mu/2$ (and super-diffusive for $\gamma > \mu/2$). The case of a finite average trapping time between two successive jumps corresponds to standard, non sub-diffusive, RWs. The fluctuations of the maximal displacement of CTRWs for $\mu = 2$ and arbitrary $\gamma < 1$ was studied in Ref. \cite{SLD2010} using real space renormalization techniques (see also \cite{CTB2010} for a Feynman-Kac approach). A precise estimate of the expected value of the maximum of the CTRW was obtained in Ref. \cite{FM2012} by combining the results of Ref. \cite{AS2005} with the so called 'subordination property'. However, nothing is known about the order statistics, beyond the first maximum, for such CTRWs. In this paper, we provide a complete analytic description of the distribution of the gap and the time interval between the first two maxima of such CTRWs for any values of the parameters $0 < \mu \leq 2$ and $0<\gamma <1$, as well as for a finite average trapping time between two successive jumps. 

We consider a CTRW starting at the origin, $x(t=0)=x_0 = 0$, and evolving according to
\begin{equation}\label{def_RW}
x_i = x_{i-1} + \eta_{i},
\end{equation}
where $x_i$ denotes the walker position between the $i$-th and the $(i+1)$-th jumps. The jumps $\eta_i$ are i.i.d. random variables distributed following a symmetric, bounded and piecewise continuous distribution $f(\eta)$ the Fourier transform of which, $\hat f(k) = \int_{-\infty}^{+\infty} f(\eta) {\rm e}^{i k \eta} d \eta$, has the small $k$ behavior
\begin{equation}\label{def_mu}
\hat f(k) = 1 - |a k|^\mu + o(|k|^\mu),
\end{equation}
where $0 < \mu \leq 2$ is the L\'evy index and $a>0$ is the characteristic length scale of the jumps. For $\mu = 2$, the variance of the jump distribution $\sigma^2 = \int_{-\infty}^{+\infty} \eta^2 f(\eta) d\eta$ is well defined and $a = \sigma /\sqrt{2}$. On the other hand, for $0 < \mu < 2$, $f(\eta)$ does not possess a well defined second moment because of its heavy tails, $f(\eta) \propto |\eta|^{-1-\mu}$ ($\eta \to \infty$), and the RW\ (\ref{def_RW}) is a L\'evy flight of index $\mu$. The time intervals $\tau_i$ between the $(i-1)$-th and the $i$-th jumps ($\tau_i>0$) are i.i.d. random variables, independent of $\eta_i$, with PDF $\Psi(\tau)$ the Laplace transform of which, $\hat\Psi(q)=\int_0^{+\infty}\Psi(\tau)\, \exp(-q\tau)\, d\tau$, has the small $q$ behavior
\begin{equation}\label{def_gamma}
\hat\Psi(q)=1-(\tau_c q)^\gamma +o(\vert q\vert^\gamma),
\end{equation}
where $0<\gamma\le 1$ and $\tau_c >0$ is the characteristic time scale of the jumps. For $\gamma =1$, the mean time between two successive jumps, $\langle\tau\rangle\equiv\int_0^{+\infty}\tau\Psi(\tau)\, d\tau <+\infty$, exists and $\tau_c=\langle\tau\rangle$. For $0<\gamma <1$, the mean time between two successive jumps does not exist. As a function of time, the walker position is thus given by
\begin{equation}\label{def_CTRW}
\frac{dx(t)}{dt}=\sum_{i=1}^n \eta_i\, \delta\left( t-\sum_{j=1}^i\tau_j\right),
\end{equation}
with $x(0)=0$ and $n$ the total number of jumps in the walk. The solution to\ (\ref{def_CTRW}) is readily found to be given by $x(\sum_{j=1}^i\tau_j <t< \sum_{j=1}^{i+1}\tau_j)=x_i$ with $x_i$ given by\ (\ref{def_RW}).

The goal of the present paper is to provide an exhaustive discussion of the different behaviors of the joint PDF of the gap $g$ and time interval $t$ between the first two maxima of the walk that may arise in the large $n$ limit, depending on the large argument behavior of $f(\eta)$ and $\Psi(\tau)$. Before entering the details of the calculations, it is useful to summarize our main results. We first show that the joint PDF $p_n(g,t)$ has a well defined limiting PDF as $n \to \infty$:
\begin{equation}\label{limit_free}
\lim_{n \to \infty} p_n(g,t) = p(g,t),
\end{equation}
where the Laplace transform of $p(g,l)$ with respect to (w.r.t.) $t$ is given in Eqs. (\ref{eq2.8}) and (\ref{eq2.5}). Then, we perform a detailed analysis of $p(g,t)$ in the plane $(g,t)$ for $\mu$ and $\gamma$ in the whole ranges $0 < \mu \leq 2$ and $0<\gamma\le 1$, and for the three different main classes identified in\ \cite{MMS2014}: (i) slow, (ii) exponentially, and (iii) fast decreasing $f(\eta)$ at large $\eta$.

One of the most remarkable result of this study may be the existence of a scaling form for the joint PDF $p(g,t)$ in the case of a fast decreasing jump distribution ($\mu =2$), such that $f(\eta + x) \sim f(\eta) \exp(-c\, \eta^{\delta}\, x)$ for $\eta \to \infty$ [see Eq. (\ref{eq5.14}) below for a more precise definition], with $0<\gamma <1$. This class includes the case of a Gaussian jump distribution for which $\delta = 1$. This is a new result, characteristic of CTRWs, without any counterpart in the discrete time RW setting considered in\ \cite{MMS2014}. More specifically, we show that there is a scaling regime $g,\ t \gg 1$ with fixed $t^{\gamma/2} g^{-\mu}$, in which $p(g,t)$ takes the scaling form
\begin{equation}\label{scaling1a}
p(g,t)\sim\frac{af(g)^2}{\tau_c (acg^\delta )^{3+2/\gamma}}
K\left(\frac{(t/\tau_c)^{\gamma/2}}{acg^\delta}\right)
\ \ \ \ (g\rightarrow +\infty\ {\rm and}\ t\rightarrow +\infty),
\end{equation}
with the asymptotic behaviors
\begin{equation}\label{scaling1b}
K(y)\sim\left\lbrace
\begin{array}{ll}
\mathcal{D}_{\rm I} \, y^{-2(1+1/\gamma)},&(y\rightarrow 0) \\
\mathcal{D}_{\rm II} \, y^{-1-2/\gamma},&(y\rightarrow +\infty)
\end{array}\right.
\end{equation}
where the amplitudes $\mathcal{D}_{\rm I}$ and $\mathcal{D}_{\rm II}$ are given in Eqs.\ (\ref{eq5.26}) and\ (\ref{eq5.27}), respectively. Physically, the switch from the first to the second behavior\ (\ref{scaling1b}) around $(t/\tau_c)^{\gamma/2}\sim acg^\delta$ corresponds to the cross-over from a `concentration' -- or `one-step' --  regime (for $(t/\tau_c)^{\gamma/2}< acg^\delta$) where the walker get stuck for a long time $t$ at the second maximum and then jumps directly to the first maximum, to a `many-steps' regime (for $(t/\tau_c)^{\gamma/2}> acg^\delta$) where she/he travels a long walk of total duration $t$ (with many steps) between the second and the first maxima. For $\gamma =1$, there is no scaling form and\ (\ref{scaling1a}) is replaced with the uniform expression\ (\ref{eq5.30}).

An other important result is how the scaling form derived in\ \cite{MMS2014} for a discrete time L\'evy flight (see Sec. 4.1 and Appendix D in\ \cite{MMS2014}) is affected when one switches to a CTRW. L\'evy flights have type (i) jump distributions since, for $0 < \mu < 2$, $f(\eta)$ necessarily has an algebraic, slow decreasing, tail. In this case we show that in the scaling regime $g,\ t \gg 1$ with fixed $t^\gamma g^{-\mu}$, the joint PDF $p(g,t)$ takes the scaling form
\begin{equation}\label{scaling2a}
p(g,t)\sim\frac{1}{a\tau_c}\, \left(\frac{a}{g}\right)^{1+\mu (1+1/\gamma)}
F_{\mu,\, \gamma}\left(\frac{a^\mu t^\gamma}{g^\mu \tau_c^\gamma}\right)
\ \ \ \ (g\rightarrow +\infty\ {\rm and}\ t\rightarrow +\infty),
\end{equation}
with the asymptotic behaviors
\begin{equation}\label{scaling2b}
F_{\mu,\, \gamma}(y)\sim\left\lbrace
\begin{array}{ll}
\mathcal{C}_{\rm I}\, y^{-1/\mu -1/\gamma},&1-{\rm Int}(\gamma) <\mu<2, \\
\mathcal{C}_{\rm II}\, y^{-1 -1/\gamma}\ln y,&\mu =1, \\
\mathcal{C}_{\rm III}\, y^{-1 -1/\gamma},&0<\mu<1,
\end{array}\right.
\ \ \ \ (y\rightarrow +\infty),
\end{equation}
and
\begin{equation}\label{scaling2c}
F_{\mu,\, \gamma}(y)\sim\mathcal{C}_{\rm IV}\, y^{1/2-1/\gamma},
\ \ \ \ (y\rightarrow 0),
\end{equation}
where ${\rm Int}(\gamma)$ in the first line of\ (\ref{scaling2b}) denotes the integer part of $\gamma$ and the last two lines are for $0<\gamma <1$ only. It can be checked that for $\gamma =1$, Eqs.\ (\ref{scaling2a}),\ (\ref{scaling2b}), and\ (\ref{scaling2c}) reduce to Eqs. (87), (88), and (90) of\ \cite{MMS2014} in which the number of jumps between the first two maxima, $l$, is merely replaced with $t/\tau_c$, as expected from simple law of large number arguments giving $l\rightarrow t/\tau_c$ as $t\rightarrow +\infty$. Our results provide thus a non trivial extension of the ones in\ \cite{MMS2014} to the case $0<\gamma <1$ in which the mean time between two successive jumps does not exist and law of large number arguments cannot be used.

Finally, by integration of the joint PDF $p(g,t)$ over $g$, one finds that the marginal distribution $p_{\rm time}(t)$ displays a power law tail with logarithmic corrections and an exponent depending only on $\mu$ and $\gamma$. Namely, for $\gamma =1$ one finds
\begin{equation}\label{eq1:result_p_of_t}
p_{\rm time}(t)\sim\frac{1}{\tau_c}\times\left\lbrace
\begin{array}{ll}
\mathcal{A}_{\rm I}\, (t/\tau_c)^{-1-1/\mu},&1<\mu\le 2, \\
\mathcal{A}_{\rm II}\, \ln(t/\tau_c)\, (t/\tau_c)^{-2},&\mu=1, \\
\mathcal{A}_{\rm III}\, (t/\tau_c)^{-2},&0<\mu<1,
\end{array}\right.
\ \ \ \ (t\rightarrow +\infty),
\end{equation}
where the amplitudes ${\cal A}_{\rm I}$, ${\cal A}_{\rm II}$ and ${\cal A}_{\rm III}$ are given in Eq.\ (\ref{eq7.1a}), and for $0<\gamma <1$ one has
\begin{equation}\label{eq2:result_p_of_t}
p_{\rm time}(t)\sim\frac{1}{\tau_c}\times\left\lbrace
\begin{array}{ll}
\mathcal{A}_{\rm I}^\prime\, (t/\tau_c)^{-1-\gamma/\mu},&1<\mu\le 2, \\
\mathcal{A}_{\rm II}^\prime\, \ln^2(t/\tau_c)\, (t/\tau_c)^{-1-\gamma},&\mu=1, \\
\mathcal{A}_{\rm III}^\prime\, \ln(t/\tau_c)\, (t/\tau_c)^{-1-\gamma},&0<\mu<1,
\end{array}\right.
\ \ \ \ (t\rightarrow +\infty),
\end{equation}
where the amplitudes ${\cal A}_{\rm I}^\prime$, ${\cal A}_{\rm II}^\prime$ and ${\cal A}_{\rm III}^\prime$ are given in Eq.\ (\ref{eq7.6a}). Again, it can be checked that Eq.\ (\ref{eq1:result_p_of_t}) is nothing but Eq. (112) in\ \cite{MMS2014} with $l=t/\tau_c$, as expected ({\it cf.} the end of the preceding paragraph). Note that for $\mu=2$ and $\gamma =1$, one has $p_{\rm time}(t) \propto t^{-3/2}$ whatever the distributions $f(\eta)$ and $\Psi(\tau)$ possessing a second and a first moment, respectively. The third lines of\ (\ref{eq1:result_p_of_t}) and\ (\ref{eq2:result_p_of_t}) reveal an interesting freezing phenomenon, as a function of $\mu$, of the large $t$ behavior of $p_{\rm time}(t)$ as $\mu$ decreases past the value $\mu_c = 1$. It follows in particular from\ (\ref{eq1:result_p_of_t}) and\ (\ref{eq2:result_p_of_t}) that the first moment of $p_{\rm time}(t)$ is never defined. This means that, although the typical size of $t$ is ${\cal O}(1)$, its average diverges with the total duration, $T$, of the random walk. For $0<\gamma\le 1$ one can estimate from\ (\ref{eq1:result_p_of_t}) and\ (\ref{eq2:result_p_of_t}) that $\langle\vert t\vert\rangle \sim T^{1-\gamma/\mu}$ for $1< \mu \le 2$, while $\langle\vert t\vert\rangle \sim T^{1-\gamma}\log{T}$ for $0 < \mu < 1$ and $\langle\vert t\vert\rangle\sim T^{1-\gamma}(\ln{T})^2$ for $\mu =1$. Table\ \ref{tab1} summarizes the functional dependence of the different asymptotic behaviors of $p_{\rm time}(t)$ at large $t$ in the $(\mu ,\gamma)$ plane.

\begin{table}[h]
\begin{tabular}{|c||c|c|c|}
\hline
$p_{\rm time}(t)\propto $&$1<\mu\le 2$&$\mu =1$&$0<\mu<1$\\
\hline\hline
$\gamma =1$&$t^{-1-1/\mu}$&$t^{-2}\ln t$&$t^{-2}$\\
\hline
$0<\gamma <1$&$t^{-1-\gamma/\mu}$&$t^{-1-\gamma}(\ln t)^2$&$t^{-1-\gamma}\ln t$ \\
\hline
\end{tabular}
\caption{ Functional dependence of the different asymptotic behaviors of $p_{\rm time}(t)$ at large $t$ in the $(\mu ,\gamma)$ plane.}
\label{tab1}
\end{table}

Note that the marginal distribution of the gap between the first two maxima, $p_{\rm gap}(g)$, obtained by integration of the joint PDF $p(g,t)$ over $t$, does not depend on $\Psi(\tau)$ (hence on $\gamma$) and is the same as the one already studied in Sec. 3 of Ref.\ \cite{MMS2014} in the discrete time RW setting.

The outline of the paper is as follows. In Section\ \ref{sec2} we give the Laplace transform of $p(g,t)$ with respect to $t$ for a free-end random walk and a random bridge in which the walker is conditioned to return to $x=0$ at the end of the walk. Section\ \ref{sec3} deals with the asymptotic behavior of $p(g,t)$ for large $t$ at fixed $g$, from which we derive, in Section\ \ref{sec4}, the asymptotic behavior of $p(g,t)$ for a L\'evy flight at large $t$ then large $g$. The asymptotic behavior of $p(g,t)$ for large $g$ is investigated in Section\ \ref{sec5} for three classes of jump distributions encompassing a wide range of jumps of practical interest. It is shown in Section\ \ref{sec5.3} that when the jump distribution has a fast decreasing tail, the behavior of $p(g,l)$ when both $g$ and $l$ are large takes on a scaling form capturing the existence of two complementary regimes, one dominated by walks with only one step between the first two maxima, and the other dominated by walks with many steps between the first two maxima. In Section\ \ref{sec6} we give the scaling form of the asymptotic behavior of $p(g,t)$ for a L\'evy flight at large $t$ and $g$. In Section\ \ref{sec7} we use the results obtained in Sections\ \ref{sec3} and\ \ref{sec6} to determine the large $t$ behavior of $p_{\rm time}(t)$ and show its freezing when the random walk is a L\'evy flight with index $0<\mu <1$. Finally, Section\ \ref{sec8} is devoted to the comparison of our analytical results with numerical simulations.
%
%%%%%%%%%%%%%%%%%%%%
%
\section{Laplace transform of $\bm{p(g,t)}$ with respect to $\bm{t}$ for a free-end random walk and a random bridge}\label{sec2}
Let $p_n(g,t)$ denote the joint PDF of the gap, $g$, and the time interval, $t$, between the first two maxima of a continuous time random walk after $n$ steps. Since $p_n(g,-t)=p_n(g,t)$ we will restrict ourselves to $t>0$ without loss of generality. Let $l$ denote the number of jumps between the first to maxima. For $t>0$ one has
\begin{equation}\label{eq2.1}
p_n(g,t)=\sum_{0<l\le n}p_n(g,l,t)=\sum_{0<l\le n}p(t\vert g,l)\, p_n(g,l)
=\sum_{0<l\le n}p(t\vert l)\, p_n(g,l),
\end{equation}
where $p(t\vert l)$ has the Laplace transform
\begin{eqnarray}\label{eq2.2}
\hat p(q\vert l)&=&\int_0^{+\infty}p(t\vert l)\, \exp(-qt)\, dt
=\int_0^{+\infty}\left\langle\delta(t-\sum_{i=1}^l \tau_i)\right\rangle\, \exp(-qt)\, dt \nonumber \\
&=&\prod_{i=1}^l \left\langle\exp(-q\tau_i)\right\rangle =\hat\Psi(q)^l .
\end{eqnarray}
Taking the Laplace transform of\ (\ref{eq2.1}) with respect to $t$ and using\ (\ref{eq2.2}) one gets
\begin{equation}\label{eq2.3}
\hat p_n(g,q)=\int_0^{+\infty}p_n(g,t)\, \exp(-qt)\, dt=
\sum_{0<l\le n}\hat\Psi(q)^l\, p_n(g,l)=\tilde p_n(g,\hat\Psi(q)),
\end{equation}
where $\tilde p_n(g,s)=\sum_{0<l\le n}p_n(g,l)\, s^l$ is the generating function of $p_n(g,l)$ with respect to $l>0$. The large $n$ limit of\ (\ref{eq2.3}) can then be readily obtained from the main result of\ \cite{MMS2014} according to which $\tilde{p}(g,s)=\lim_{n\rightarrow +\infty}\tilde{p}_n(g,s)$ exists and is given by
\begin{equation}\label{eq2.4}
\tilde{p}(g,s)=I_1(g,s)I_2(g),
\end{equation}
with
\begin{equation}\label{eq2.5}
\begin{array}{l}
I_1(g,s)=s\int_0^{+\infty}u(y,s) f(g+y)\, dy, \\
I_2(g)=\int_0^{+\infty}h(x,1) f(g+x)\, dx,
\end{array}
\end{equation}
where the functions $u(x,s)$ and $h(x,s)$ are the inverse Laplace transforms of $\phi(\lambda ,s)$ and $\phi(\lambda ,s)/\lambda$, respectively:
\begin{equation}\label{eq2.6}
\begin{array}{l}
\int_0^{+\infty}u(x,s){\rm e}^{-\lambda x}dx=\phi(\lambda ,s), \\
\int_0^{+\infty}h(x,s){\rm e}^{-\lambda x}dx=\phi(\lambda ,s)/\lambda ,
\end{array}
\end{equation}
with
\begin{equation}\label{eq2.7}
\phi(\lambda ,s)=\exp\left(-\frac{\lambda}{\pi}
\int_0^{+\infty}\frac{\ln\lbrack 1-s\hat{f}(k)\rbrack}{k^2+\lambda^2}\, dk\right) .
\end{equation}
Injecting this result into the large $n$ limit of\ (\ref{eq2.3}) yields
\begin{equation}\label{eq2.8}
\hat p(g,q)=\lim_{n\rightarrow +\infty}\hat p_n(g,q)=
\tilde p(g,\hat\Psi(q))=I_1(g,\hat\Psi(q))I_2(g).
\end{equation}
Thus, $p_n(g,t)$ converges to a limiting distribution $p(g,t)$ as $n\rightarrow +\infty$ whose Laplace transform with respect to $t$, $\hat p(g,q)=\int_0^{+\infty}p(g,t)\, \exp(-qt)\, dt$, is given by\ (\ref{eq2.8}).

It was proved in\ \cite{MMS2014} that the expression\ (\ref{eq2.4}) of $\tilde{p}(g,s)$ actually holds for both free-end random walks and random bridges in which the walker is conditioned to return to $x=0$ at the end of the walk. Since, for a given $\Psi(\tau)$, $\hat p(g,q)$ is entirely determined by $\tilde{p}(g,s)$ through\ (\ref{eq2.8}), it follows immediately that the expression of $\hat p(g,q)$, hence the one of the limiting joint PDF $p(g,t)$, is exactly the same for free-end walks and bridges. Therefore, all the results obtained in the following apply to free-end random walks as well as to random bridges without any modification.
%
%%%%%%%%%%%%%%%%%%%%
%
\section{Asymptotic behavior of $\bm{p(g,t)}$ for large $\bm{t}$ at fixed $\bm{g}$}\label{sec3}
Sections\ \ref{sec3} to\ \ref{sec6} are devoted to the asymptotic behavior of the joint PDF $p(g,t)$ when either $g$ or $t$ (or both) is large. From the results obtained for the large $t$ behavior of $p(g,t)$ it will then be possible to derive the large $t$ behavior of the marginal distribution $p_{\rm time}(t)=\int_0^{+\infty}p(g,t)\, dg$, which will be the subject of Sec.\ \ref{sec7}.

First, we consider the limit $t\rightarrow +\infty$ at fixed $g$. As we will see, the behavior of $p(g,t)$ in this limit depends on the value of $\gamma$, leading to new, non trivial, asymptotic expressions if $0<\gamma <1$.
\subsection{$\bm{\gamma =1}$}\label{sec3.1}
In this case the mean time between two successive jumps, $\langle\tau\rangle =\tau_c$, exists and by a simple law of large number argument, the large $t$ behavior of $p(g,t)$ can be expected to be given by $\tau_c^{-1}p(g,l=t/\tau_c)$, where $l=t/\tau_c$ is large. Let us check this result explicitly by comparing the large $t$ behavior of $p(g,t)$ and the large $l$ behavior of $p(g,l)$.

The behavior of $p(g,t)$ at large $t$ is determined by the one of $\hat p(g,q)$ at small $q$. Thus, from\ (\ref{eq2.8}) and\ (\ref{def_gamma}) with $\gamma =1$, one has
\begin{eqnarray}\label{eq3.1}
p(g,t)&=&\frac{1}{2i\pi}\int_{\mathcal L}\tilde p(g,\hat\Psi(q))\, \exp(qt)\, dq \nonumber \\
&\sim&\frac{1}{2i\pi}\int_{\mathcal L}\tilde p(g,1-\tau_c q)\, \exp(qt)\, dq
\ \ \ \ (t\rightarrow +\infty) \nonumber \\
&=&\frac{1}{2i\pi\, \tau_c}\int_{\mathcal L}\tilde p(g,1-q)\, \exp\left(\frac{qt}{\tau_c}\right)\, dq,
\end{eqnarray}
where we have made the change of variable $q\rightarrow q/\tau_c$. Similarly, for large $l$ the behavior of $p(g,l)$ is determined by the one of $\tilde p(g,s)$ near $s=1$. Making the change of variable $s=\exp(-q)$ one gets
\begin{eqnarray}\label{eq3.2}
p(g,l)&=&\frac{1}{2i\pi}\oint\frac{\tilde p(g,s)}{s^{l+1}}ds \nonumber \\
&=&\frac{1}{2i\pi}\int_{-i\pi +0}^{+i\pi +0}\tilde p(g,\exp(-q))\, \exp(ql)\, dq \nonumber \\
&\sim&\frac{1}{2i\pi}\int_{\mathcal L}\tilde p(g,1-q)\, \exp(ql)\, dq
\ \ \ \ (l\rightarrow +\infty).
\end{eqnarray}
Comparing Eqs.\ (\ref{eq3.1}) and\ (\ref{eq3.2}), one obtains the expected expression
\begin{equation}\label{eq3.3}
p(g,t)\sim\frac{1}{\tau_c}p\left(g,l=\frac{t}{\tau_c}\right)\ \ \ \ (t\rightarrow +\infty).
\end{equation}
Finally, Eq.\ (65) of\ \cite{MMS2014} together with Eq.\ (\ref{eq3.3}) above yield, for all $0<\mu\le 2$,
\begin{equation}\label{eq3.4}
p(g,t)\sim\frac{\Gamma(1+1/\mu)}{\pi a\tau_c}\, \frac{I_2(g)^2}{(t/\tau_c)^{1+1/\mu}}
\ \ \ \ (t\rightarrow +\infty).
\end{equation}
%
%%%%%%%%%%%%%%%%%%%%
%
\subsection{$\bm{0<\gamma <1}$}\label{sec3.2}
If $0<\gamma <1$ the mean time between two successive jumps, $\langle\tau\rangle =+\infty$, does not exist and the large $t$ behavior of $p(g,t)$ is no longer related to $p(g,l)$ by the simple expression\ (\ref{eq3.3}). Following the same line as for Eq.\ (\ref{eq3.1}) with $0<\gamma <1$, one now has
\begin{eqnarray}\label{eq3.5}
p(g,t)&=&\frac{1}{2i\pi}\int_{\mathcal L}\tilde p(g,\hat\Psi(q))\, \exp(qt)\, dq \nonumber \\
&\sim&\frac{1}{2i\pi}\int_{\mathcal L}\tilde p(g,1-(\tau_c q)^\gamma)\, \exp(qt)\, dq
\ \ \ \ (t\rightarrow +\infty) \nonumber \\
&=&\frac{1}{2i\pi\, \tau_c}\int_{\mathcal L}\tilde p(g,1-q^\gamma)\, \exp\left(\frac{qt}{\tau_c}\right)\, dq,
\end{eqnarray}
where we have made the change of variable $q\rightarrow q/\tau_c$. It remains to compute the inverse Laplace transform on the right-hand side of\ (\ref{eq3.5}). One finds three different regimes depending on the value of $\mu$.
%
%%%%%%%%%%%%%%%%%%%%
%
\subsubsection{$\bm{1<\mu\le 2}$}\label{sec3.2.1}
For $1<\mu\le 2$ the behavior of $\tilde p(g,s)$ near $s=1$ is given by, (see Eq.\ (63) in\ \cite{MMS2014}),
\begin{equation}\label{eq3.6}
\tilde p(g,s)-\tilde p(g,1)\sim -a_\mu I_2(g)^2 (1-s)^{1/\mu}\ \ \ \ (s\rightarrow 1),
\end{equation}
with $a_\mu=1/\lbrack a\sin(\pi/\mu)\rbrack$. Injecting\ (\ref{eq3.6}) into\ (\ref{eq3.5}) yields
\begin{equation}\label{eq3.7}
p(g,t)\sim -\frac{a_\mu I_2(g)^2}{2i\pi\, \tau_c}
\int_{\mathcal L} q^{\gamma/\mu}\, \exp\left(\frac{qt}{\tau_c}\right)\, dq
\ \ \ \ (t\rightarrow +\infty).
\end{equation}
Performing then the inverse Laplace transform in\ (\ref{eq3.7}) and using the reflection formula $\Gamma(-z)\Gamma(z+1)=-\pi/\sin(\pi z)$ one obtains, for all $1<\mu \le 2$,
\begin{equation}\label{eq3.8}
p(g,t)\sim\frac{\sin(\pi\gamma /\mu)}{\sin(\pi /\mu)}\, 
\frac{\Gamma(1+\gamma /\mu)}{\pi a\tau_c}\, \frac{I_2(g)^2}{(t/\tau_c)^{1+\gamma /\mu}}
\ \ \ \ (t\rightarrow +\infty).
\end{equation}
Note that the exponent of $t$ in\ (\ref{eq3.8}) depends on $\mu$ and\ (\ref{eq3.8}) tends to\ (\ref{eq3.4}) as $\gamma\rightarrow 1$.
%
%%%%%%%%%%%%%%%%%%%%
%
\subsubsection{$\bm{0<\mu <1}$}\label{sec3.2.2}
For $0<\mu <1$ the behavior of $\tilde p(g,s)$ near $s=1$ reads, (see Eqs.\ (63) and\ (C9) in\ \cite{MMS2014}),
\begin{equation}\label{eq3.9}
\tilde p(g,s)-\tilde p(g,1)\sim -I_2(g)J_1(g) (1-s)\ \ \ \ (s\rightarrow 1),
\end{equation}
with
\begin{equation}\label{eq3.10}
J_1(g)=\int_0^{+\infty}w_1(x)\, f(g+x)\, dx,
\end{equation}
where $w_1(x)$ is defined by its Laplace transform
\begin{equation}\label{eq3.11}
\int_0^{+\infty}w_1(x){\rm e}^{-\lambda x}dx=\frac{\lambda\phi(\lambda ,1)}{\pi}
\int_0^{+\infty}\frac{\hat F(k)}{k^2+\lambda^2}\, dk,
\end{equation}
with $\hat F(k)=\hat f(k)/(1-\hat f(k))$. Equations\ (\ref{eq3.5}) and\ (\ref{eq3.9}) now yield
\begin{equation}\label{eq3.12}
p(g,t)\sim -\frac{I_2(g)J_1(g)}{2i\pi\, \tau_c}
\int_{\mathcal L} q^{\gamma}\, \exp\left(\frac{qt}{\tau_c}\right)\, dq
\ \ \ \ (t\rightarrow +\infty),
\end{equation}
which replaces Eq.\ (\ref{eq3.8}). Performing the inverse Laplace transform in\ (\ref{eq3.12}) and using the reflection formula $\Gamma(-z)\Gamma(z+1)=-\pi/\sin(\pi z)$ one obtains, for all $0<\mu < 1$,
\begin{equation}\label{eq3.13}
p(g,t)\sim\frac{\sin(\pi\gamma)\, \Gamma(1+\gamma)}{\pi\tau_c}\, 
\frac{I_2(g)J_1(g)}{(t/\tau_c)^{1+\gamma}}
\ \ \ \ (t\rightarrow +\infty).
\end{equation}
Note that the exponent of $t$ in\ (\ref{eq3.13}) does {\it not} depend on $\mu$, unlike its counterpart in Eq.\ (\ref{eq3.8}). It follows in particular that\ (\ref{eq3.13}) does not tend to\ (\ref{eq3.4}) as $\gamma\rightarrow 1$. This exponent `freezing' from $1+\gamma/\mu$ to $1+\gamma$ as $\mu$ decreases past the critical value $\mu_c =1$ (at fixed $0<\gamma <1$) is due to a switch of leading terms in the asymptotic expansion of $\tilde p(g,s)-\tilde p(g,1)$ near $s=1$ from $\propto (1-s)^{1/\mu}$ for $1<\mu\le 2$, to $\propto (1-s)$ for $0<\mu <1$. (See Eqs.\ (63) and\ (C9) in\ \cite{MMS2014}). Although a term $\propto (1-s)$, as regular, does not contribute to the large $l$ behavior of $p(g,l)$, it gives a term $\propto q^\gamma$ in the asymptotic expansion of $\tilde p(g,1-q^\gamma)$ near $q=0$ in\ (\ref{eq3.5}) which is singular if $0<\gamma <1$ and thus contributes to (and dominates) the large $t$ behavior of $p(g,t)$ for $0<\mu <1$.
%
%%%%%%%%%%%%%%%%%%%%
%
\subsubsection{$\bm{\mu =1}$ \bf{and summary}}\label{sec3.2.3}
Finally, at the critical value $\mu =\mu_c =1$, the equation (C9) in\ \cite{MMS2014} gives
\begin{equation}\label{eq3.14}
\tilde p(g,s)-\tilde p(g,1)\sim -\frac{I_2(g)^2}{\pi a}\, (1-s)\ln\left(\frac{1}{1-s}\right)\ \ \ \ (s\rightarrow 1),
\end{equation}
which, together with Eq.\ (\ref{eq3.5}), yields
\begin{equation}\label{eq3.15}
p(g,t)\sim \frac{\gamma I_2(g)^2}{2i\pi^2 a\tau_c}
\int_{\mathcal L} q^{\gamma}\ln(q)\, \exp\left(\frac{qt}{\tau_c}\right)\, dq
\ \ \ \ (t\rightarrow +\infty).
\end{equation}
Using then
\begin{equation*}
\frac{1}{2i\pi}\int_{\mathcal L} q^{\gamma}\ln(q)\, \exp(qt)\, dq =
-\frac{1}{\Gamma(-\gamma)}\, \frac{\ln(t)}{t^{1+\gamma}}
+\frac{\Gamma^\prime(-\gamma)}{\Gamma(-\gamma)^2}\, \frac{1}{t^{1+\gamma}},
\end{equation*}
and the reflection formula $\Gamma(-z)\Gamma(z+1)=-\pi/\sin(\pi z)$, one obtains
\begin{equation}\label{eq3.16}
p(g,t)\sim\frac{\gamma\sin(\pi\gamma)\, \Gamma(1+\gamma)}{\pi^2 a\tau_c}\, 
\frac{I_2(g)^2 \ln(t/\tau_c)}{(t/\tau_c)^{1+\gamma}}
\ \ \ \ (t\rightarrow +\infty).
\end{equation}

Defining
\begin{equation}\label{eq3.17a}
\mathcal{B}_{\rm I}=\frac{\sin(\pi\gamma/\mu)}{\pi\sin(\pi/\mu)}\, 
\Gamma(1+\gamma/\mu),
\end{equation}
and
\begin{equation}\label{eq3.17b}
\mathcal{B}_{\rm II}=\frac{\sin(\pi\gamma)}{\pi^2}\, 
\Gamma(1+\gamma),
\end{equation}
we can gather the equations\ (\ref{eq3.4}),\ (\ref{eq3.8}),\ (\ref{eq3.13}), and\ (\ref{eq3.16}) in a more concise form
\begin{equation}\label{eq3.17}
p(g,t)\sim\frac{1}{a\tau_c}\times\left\lbrace
\begin{array}{ll}
\mathcal{B}_{\rm I}\, I_2(g)^2 (\tau_c/t)^{1+\gamma/\mu},&1-{\rm Int}(\gamma) <\mu\le 2, \\
\gamma\mathcal{B}_{\rm II}\, I_2(g)^2 (\tau_c/t)^{1+\gamma}\ln(t/\tau_c),&\mu=1, \\
\pi a\mathcal{B}_{\rm II}\, I_2(g)J_1(g) (\tau_c/t)^{1+\gamma},&0<\mu<1,
\end{array}
\right.\ \ \ \ 
\begin{array}{r}(t\rightarrow +\infty\\ {\rm at\ fixed\ } g)\end{array}
\end{equation}
which summarizes the large $t$ behavior of $p(g,t)$ at fixed $g$ and $0<\gamma\le 1$. The term ${\rm Int}(\gamma)$ in the first line of\ (\ref{eq3.17}) denotes the integer part of $\gamma$ and the last two lines are for $0<\gamma <1$ only.
%
%%%%%%%%%%%%%%%%%%%%
%
\section{Asymptotic behavior of $\bm{p(g,t)}$ for a L\'evy flight at large $\bm{t}$ then large $\bm{g}$}\label{sec4}
The random walk is a L\'evy flight if $0<\mu <2$, in which case the variance of $f(\eta)$ does not exist and $f(\eta)\sim\eta^{-1-\mu}$ at large $\eta$. In this section we investigate the asymptotic behavior of $p(g,t)$ for a L\'evy flight in the limit $t\rightarrow +\infty$ then $g\rightarrow +\infty$.

In the cases $\gamma =1$ with $0<\mu <2$ and $0<\gamma <1$ with $1\le\mu <2$, this asymptotic behavior follows readily from Eqs.\ (\ref{eq3.4}),\ (\ref{eq3.8}),\ (\ref{eq3.16}), and the large $g$ behavior of $I_2(g)$:
\begin{equation}\label{eq4.1}
I_2(g)\sim\frac{a^{\mu/2}}{\Gamma(1-\mu/2)}\, \frac{1}{g^{\mu/2}}\ \ \ \ (g\rightarrow +\infty),
\end{equation}
(see Eq.\ (48) in \cite{MMS2014}). One finds,
\begin{equation}\label{eq4.2}
p(g,t)\sim\frac{1}{a\tau_c}\, \frac{\mathcal{B}_{\rm I}}{\Gamma(1-\mu/2)^2}\, 
\left(\frac{a}{g}\right)^\mu \left(\frac{\tau_c}{t}\right)^{1+\gamma/\mu}
\ \ \ \ (t\rightarrow +\infty\ {\rm then}\ g\rightarrow +\infty),
\end{equation}
for $\gamma =1$ with $0<\mu <2$ and $0<\gamma <1$ with $1<\mu <2$, and
\begin{equation}\label{eq4.3}
p(g,t)\sim\frac{1}{a\tau_c}\, \frac{\gamma\mathcal{B}_{\rm II}}{\pi}\, 
\left(\frac{a}{g}\right)  \left(\frac{\tau_c}{t}\right)^{1+\gamma}\ln\left(\frac{t}{\tau_c}\right)
\ \ \ \ (t\rightarrow +\infty\ {\rm then}\ g\rightarrow +\infty),
\end{equation}
for $0<\gamma <1$ with $\mu =1$.

It remains the case $0<\gamma <1$ with $0<\mu <1$. Injecting the asymptotics
\begin{equation}\label{eq4.4}
\phi(\lambda ,1)\sim\frac{1}{(a\lambda)^{\mu/2}}\ \ \ \ (\lambda\rightarrow 0),
\end{equation}
(see Eq.\ (41) in\ \cite{MMS2014}), and
\begin{eqnarray}\label{eq4.5}
\int_0^{+\infty}\frac{\lambda\, \hat{F}(k)}{k^2+\lambda^2}\, dk
&\sim&\frac{1}{(a\lambda)^\mu}\int_0^{+\infty}\frac{dq}{q^\mu (1+q^2)}
\ \ \ \ (\lambda\rightarrow 0) \nonumber \\
&=&\frac{\pi}{2\cos(\mu\pi /2)}\, \frac{1}{(a\lambda)^\mu},
\end{eqnarray}
into the right-hand side of\ (\ref{eq3.11}), one gets
\begin{equation}\label{eq4.6}
\int_0^{+\infty}w_1(x){\rm e}^{-\lambda x}dx\sim
\frac{1}{2\cos(\mu\pi /2)}\, \frac{1}{(a\lambda)^{3\mu/2}}\ \ \ \ (\lambda\rightarrow 0),
\end{equation}
the inverse Laplace transform of which gives
\begin{equation}\label{eq4.7}
w_1(x)\sim\frac{x^{3\mu/2 -1}}{2a^{3\mu/2}\cos(\mu\pi/2)\Gamma(3\mu/2)}
\ \ \ \ (x\rightarrow +\infty).
\end{equation}
To determine the large $g$ behavior of $J_1(g)$ we make the change of variable $x=g\overline x$ in Eq.\ (\ref{eq3.10}), let $g\rightarrow +\infty$, and use\ (\ref{eq4.7}) and the large $x$ behavior of $f(x)$:
\begin{equation}\label{eq4.8}
f(x)\sim\sin\left(\frac{\mu\pi}{2}\right)\, \Gamma(\mu +1)\frac{a^\mu}{\pi x^{\mu +1}}
\ \ \ \ (x\rightarrow +\infty),
\end{equation}
(see Eq.\ (46) in\ \cite{MMS2014}). One finds
\begin{eqnarray}\label{eq4.9}
J_1(g)&=&
g\int_0^{+\infty}w_1(g\overline x)f\lbrack g(1+\overline x)\rbrack\, d\overline x \nonumber \\
&\sim&\frac{\tan(\mu\pi/2)}{2\pi\, a^{\mu/2}}\, \frac{\Gamma(\mu +1)}{\Gamma(3\mu/2)}\,
\frac{1}{g^{1-\mu/2}}
\int_0^{+\infty}\frac{\overline{x}^{3\mu/2-1}}{(1+\overline{x})^{\mu +1}}\, d\overline{x}
\ \ \ \ (g\rightarrow +\infty) \nonumber \\
&=&\frac{\tan(\mu\pi/2)\Gamma(1-\mu/2)}{2\pi\, a^{\mu/2}}\, \frac{1}{g^{1-\mu/2}},
\end{eqnarray}
where the integral over $\overline x$ in the second line is equal to $\Gamma(1-\mu/2)\Gamma(3\mu/2)/\Gamma(\mu +1)$. Thus, from\ (\ref{eq3.13}),\ (\ref{eq4.1}), and\ (\ref{eq4.9}) one obtains
\begin{equation}\label{eq4.10}
p(g,t)\sim\frac{1}{a\tau_c}\, \frac{\mathcal{B}_{\rm II}\tan(\mu\pi/2)}{2}\, 
\left(\frac{a}{g}\right)  \left(\frac{\tau_c}{t}\right)^{1+\gamma}
\ \ \ \ (t\rightarrow +\infty\ {\rm then}\ g\rightarrow +\infty).
\end{equation}

Finally, writing $\mathcal{C}_{\rm I}=\mathcal{B}_{\rm I}\Gamma(1-\mu/2)^{-2}$, $\mathcal{C}_{\rm II}=\mathcal{B}_{\rm II}\pi^{-1}$, and $\mathcal{C}_{\rm III}=\mathcal{B}_{\rm II}\tan(\mu\pi/2)/2$, one has
\begin{equation}\label{eq4.11}
p(g,t)\sim\frac{1}{a\tau_c}\times\left\lbrace
\begin{array}{ll}
\mathcal{C}_{\rm I}\, (a/g)^\mu (\tau_c/t)^{1+\gamma/\mu},&1-{\rm Int}(\gamma) <\mu<2, \\
\gamma\mathcal{C}_{\rm II}\, (a/g)\, (\tau_c/t)^{1+\gamma}\ln(t/\tau_c),&\mu=1, \\
\mathcal{C}_{\rm III}\, (a/g)\, (\tau_c/t)^{1+\gamma},&0<\mu<1,
\end{array}
\right.\ \ \ \ 
\begin{array}{r}(t\rightarrow +\infty\ {\rm then}\\ g\rightarrow +\infty)\end{array}
\end{equation}
where ${\rm Int}(\gamma)$ in the first line denotes the integer part of $\gamma$ and the last two lines are for $0<\gamma <1$ only.

%
%%%%%%%%%%%%%%%%%%%%
%
\section{Asymptotic behavior of $\bm{p(g,t)}$ for large $\bm{g}$}\label{sec5}
Now, we consider the asymptotic behavior of $p(g,t)$ in the large $g$ limit. If the support of the jump distribution $f(\eta)$ is bounded, the first gap cannot be larger than the (finite) diameter of this support and one trivially has $p(g,t)=0$ for every $t$ and $g>g_0$ with $g_0<+\infty$ large enough. If the support of $f(\eta)$ is not bounded the result depends on the tail of $f(\eta)$. We have singled out three classes of tails which encompass a wide range of jumps of practical interest.
%
%%%%%%%%%%%%%%%%%%%%
%
\subsection{Slow decreasing $\bm{f(\eta)}$}\label{sec5.1}
Slow decreasing jump distributions are defined by
\begin{equation}\label{eq5.1}
f(g+x)\sim f(g) \ \ \ \ (g\rightarrow +\infty),
\end{equation}
for any fixed $x$. Jump distributions with an algebraic tail are slow decreasing. The large $g$ behavior of $I_1(g,s)$ is then readily obtained from\ (\ref{eq2.5}),\ (\ref{eq2.6}), and\ (\ref{eq5.1}). One gets
\begin{equation}\label{eq5.2}
I_1(g,s)\sim sf(g)\int_0^{+\infty}u(y,s)\, dy=sf(g)\phi(0,s)
=\frac{sf(g)}{\sqrt{1-s}} \ \ \ \ (g\rightarrow +\infty),
\end{equation}
where we have used $\phi(0,s)=1/\sqrt{1-s}$ as given by Eq.\ (\ref{eq2.7}) in the limit $\lambda\rightarrow 0$, (see also e.g.\ \cite{AS2005}). The large $g$ behavior of $I_2(g)$ depends on $f(\eta)$ and cannot be specified further on at this point. Thus, Eqs\ (\ref{eq2.4}) and\ (\ref{eq5.2}) yield
\begin{equation}\label{eq5.3}
\tilde{p}(g,s)\sim\frac{sf(g)I_2(g)}{\sqrt{1-s}} \ \ \ \ (g\rightarrow +\infty),
\end{equation}
and from Eqs.\ (\ref{eq2.8}) and\ (\ref{eq5.3}) one finds
\begin{eqnarray}\label{eq5.4}
p(g,t)&=&\frac{1}{2i\pi}\int_{\mathcal L}\tilde p(g,\hat\Psi(q))\, \exp(qt)\, dq \nonumber \\
&\sim&\frac{f(g)I_2(g)}{2i\pi}\int_{\mathcal L}
\frac{\hat\Psi(q)}{\sqrt{1-\hat\Psi(q)}}\, \exp(qt)\, dq
\ \ \ \ (g\rightarrow +\infty).
\end{eqnarray}
The large $t$ behavior of\ (\ref{eq5.4}) is then obtained from the small $q$ behavior of $\hat\Psi(q)$, Eq.\ (\ref{def_gamma}). One has
\begin{eqnarray}\label{eq5.5}
\frac{1}{2i\pi}\int_{\mathcal L}\frac{\hat\Psi(q)}{\sqrt{1-\hat\Psi(q)}}\, \exp(qt)\, dq
&\sim&\frac{1}{2i\pi\, \tau_c^{\gamma/2}}\int_{\mathcal L}\frac{\exp(qt)}{q^{\gamma/2}}\, dq
\ \ \ \ (t\rightarrow +\infty) \nonumber \\
&=&\frac{1}{\Gamma(\gamma/2)\tau_c}\, \frac{1}{(t/\tau_c)^{1-\gamma/2}},
\end{eqnarray}
which together with Eq.\ (\ref{eq5.4}) gives
\begin{equation}\label{eq5.6}
p(g,t)\sim\frac{1}{\Gamma(\gamma/2)\tau_c}\, \frac{f(g)I_2(g)}{(t/\tau_c)^{1-\gamma/2}}
\ \ \ \ (g\rightarrow +\infty\ {\rm then}\ t\rightarrow +\infty).
\end{equation}

For a L\'evy flight of index $0<\mu <2$, the large $g$ behaviors of $f(g)$ and $I_2(g)$ are respectively given by\ (\ref{eq4.8}) and\ (\ref{eq4.1}), and Eqs.\ (\ref{eq5.4}) and\ (\ref{eq5.6}) read
\begin{equation}\label{eq5.4a}
p(g,t)\sim \frac{\mathcal{C}_{\rm IV}\Gamma(\gamma/2)}{a}
\, \left(\frac{a}{g}\right)^{1+3\mu/2}\Xi(t)
\ \ \ \ (g\rightarrow +\infty),
\end{equation}
and
\begin{equation}\label{eq5.6a}
p(g,t)\sim\frac{\mathcal{C}_{\rm IV}}{a\tau_c}\, \left(\frac{a}{g}\right)^{1+3\mu/2}
\left(\frac{\tau_c}{t}\right)^{1-\gamma/2}\ \ \ \ (g\rightarrow +\infty\ {\rm then}\ t\rightarrow +\infty),
\end{equation}
respectively, with
\begin{equation}\label{eq5.6b}
\mathcal{C}_{\rm IV}=\sin(\mu\pi/2)\, \frac{\Gamma(\mu +1)}
{\pi\Gamma(\gamma/2)\Gamma(1-\mu/2)},
\end{equation}
and
\begin{equation}\label{eq5.6c}
\Xi(t)=\frac{1}{2i\pi}\int_{\mathcal L}\frac{\hat\Psi(q)}{\sqrt{1-\hat\Psi(q)}}\, \exp(qt)\, dq.
\end{equation}
%
%%%%%%%%%%%%%%%%%%%%
%
\subsection{Exponentially decreasing $\bm{f(\eta)}$}\label{sec5.2}
Exponentially decreasing jump distributions are defined by
\begin{equation}\label{eq5.7}
f(g+x)\sim f(g)\exp(-cx) \ \ \ \ (g\rightarrow +\infty),
\end{equation}
for some $c>0$ and any fixed $x$. Substituting\ (\ref{eq5.7}) for $f(g+x)$ on the right-hand side of\ (\ref{eq2.5}) and using\ (\ref{eq2.6}), one gets
\begin{equation}\label{eq5.8}
I_1(g,s)\sim f(g)\, s\phi(c,s) \ \ \ \ (g\rightarrow +\infty),
\end{equation}
and
\begin{equation}\label{eq5.9}
I_2(g)\sim f(g)\, \frac{\phi(c,1)}{c} \ \ \ \ (g\rightarrow +\infty),
\end{equation}
which, together with Eq.\ (\ref{eq2.4}), yield
\begin{equation}\label{eq5.10}
\tilde{p}(g,s)\sim \frac{\phi(c,1)}{c}\, f(g)^2\, s\phi(c,s) \ \ \ \ (g\rightarrow +\infty),
\end{equation}
and from Eqs.\ (\ref{eq2.8}) and\ (\ref{eq5.10}) one finds
\begin{eqnarray}\label{eq5.11}
p(g,t)&=&\frac{1}{2i\pi}\int_{\mathcal L}\tilde p(g,\hat\Psi(q))\, \exp(qt)\, dq \nonumber \\
&\sim&\frac{\phi(c,1)\, f(g)^2}{2i\pi c}\int_{\mathcal L}
\hat\Psi(q)\phi(c,\hat\Psi(q))\, \exp(qt)\, dq
\ \ \ \ (g\rightarrow +\infty).
\end{eqnarray}
The large $t$ behavior of\ (\ref{eq5.11}) is obtained from the behavior of\ (\ref{eq5.10}) near its dominant singularity at $s=1$ and the small $q$ behavior of $\hat\Psi(q)$, Eq.\ (\ref{def_gamma}). The former is given by, (see Eq.\ (60) in\ \cite{MMS2014} with $\mu =2$ and $\lambda =c$),
\begin{equation}\label{eq5.12}
\tilde{p}(g,s)-\tilde{p}(g,1)\sim -\frac{1}{a}\left\lbrack\frac{\phi(c,1)}{c}\right\rbrack^2
f(g)^2 s\sqrt{1-s} \ \ \ \ (g\rightarrow +\infty\ {\rm then}\ s\rightarrow 1).
\end{equation}
Injecting then\ (\ref{eq5.12}) into the right-hand side of\ (\ref{eq5.11}) and performing the inverse Laplace transform with $\hat\Psi(q)$ given by\ (\ref{def_gamma}), one finds
\begin{eqnarray}\label{eq5.13}
p(g,t)&\sim&-\frac{1}{a}\left\lbrack\frac{\phi(c,1)}{c}\right\rbrack^2
\frac{f(g)^2}{2i\pi}\int_{\mathcal L}(\tau_c q)^{\gamma/2}\, \exp(qt)\, dq
\ \ \ \ (g\rightarrow +\infty\ {\rm then}\ t\rightarrow +\infty) \nonumber \\
&=&\frac{\sin(\pi\gamma/2)\Gamma(1+\gamma/2)}{\pi a\tau_c}\left\lbrack\frac{\phi(c,1)}{c}\right\rbrack^2\frac{f(g)^2}{(t/\tau_c)^{1+\gamma/2}} ,
\end{eqnarray}
where we have used the reflection formula $\Gamma(-z)\Gamma(z+1)=-\pi/\sin(\pi z)$.
%
%%%%%%%%%%%%%%%%%%%%
%
\subsection{Fast decreasing $\bm{f(\eta)}$}\label{sec5.3}
Fast decreasing jump distributions are defined by
\begin{equation}\label{eq5.14}
f(g+x)\sim f(g)\exp(-cg^\delta x+\theta(x,g)) \ \ \ \ (g\rightarrow +\infty),
\end{equation}
for some $c,\delta>0$, any fixed $x$, and where $\theta(x,g)$ is such that $\lim_{g\rightarrow +\infty}\theta(x\lesssim g^{-\delta},g)=0$. Super-exponentially distributed jumps are fast decreasing. Substituting\ (\ref{eq5.14}) for $f(g+x)$ on the right-hand side of\ (\ref{eq2.5}) and using\ (\ref{eq2.6}), one gets, (see Sec. 4.3 in\ \cite{MMS2014} for details),
\begin{eqnarray}\label{eq5.15}
I_1(g,s)&\sim& f(g)s\phi(cg^\delta ,s) \nonumber \\
&\sim& f(g)\, \left(
s-\frac{s}{\pi cg^\delta}\int_0^{+\infty}\ln\lbrack 1-s\hat{f}(k)\rbrack\, dk\right)
 \ \ \ \ (g\rightarrow +\infty),
\end{eqnarray}
and
\begin{equation}\label{eq5.16}
I_2(g)\sim f(g)\frac{\phi(cg^\delta ,1)}{cg^\delta}
\sim\frac{f(g)}{cg^\delta} \ \ \ \ (g\rightarrow +\infty).
\end{equation}
Eqs.\ (\ref{eq5.15}) and\ (\ref{eq5.16}), together with Eq.\ (\ref{eq2.4}), yield
\begin{equation}\label{eq5.17}
\tilde{p}(g,s)\sim\frac{f(g)^2}{cg^\delta}\, \left(
s-\frac{s}{\pi cg^\delta}\int_0^{+\infty}\ln\lbrack 1-s\hat{f}(k)\rbrack\, dk\right)
 \ \ \ \ (g\rightarrow +\infty),
\end{equation}
from which one readily obtains $p(g,l)$ for large $g$, by expanding the logarithm in power series of $s$, (see Eq.\ (103) in\ \cite{MMS2014}). One finds in particular,
\begin{equation}\label{eq5.18}
p(g,l=1)\sim\frac{f(g)^2}{cg^{\delta}} \ \ \ \ (g\rightarrow +\infty),
\end{equation}
and
\begin{equation}\label{eq5.19}
\frac{p(g,l)}{p(g,1)}\sim\frac{\alpha_l}{g^{\delta}} \underset{g \rightarrow\infty}{\rightarrow} 0,
\end{equation}
for all $l>1$, where $\alpha_l$ depends on $l$ only. These results reveal a concentration of $p(g,l)$ onto $l=\pm 1$ for fast decreasing jump distributions as $g\rightarrow +\infty$. From this concentration and the $n\rightarrow +\infty$ limit of\ (\ref{eq2.1}), $p(g,t)=\sum_{0<l\le n}p(t\vert l)\, p(g,l)$, one has
\begin{equation}\label{eq5.20}
p(g,t)\sim p(g,l=1)p(t\vert l=1)\ \ \ \ (g\rightarrow +\infty),
\end{equation}
which reads, using $p(t\vert l=1)=\Psi(t)$ and Eq.\ (\ref{eq5.18}),
\begin{equation}\label{eq5.21}
p(g,t)\sim\frac{f(g)^2 \Psi(t)}{c g^\delta}\ \ \ \ (g\rightarrow +\infty).
\end{equation}
For $\gamma <1$, $\Psi(t)$ has an algebraic tail at large $t$, $\Psi(t)\sim -\Gamma(-\gamma)\tau_c^\gamma t^{-1-\gamma}$, which follows from the small $q$ behavior\ (\ref{def_gamma}) of $\hat\Psi(q)$, and one gets the large $t$ behavior of\ (\ref{eq5.21}) as
\begin{equation}\label{eq5.22}
p(g,t)\sim\frac{\sin(\pi\gamma)\Gamma(1+\gamma)}{\pi c\tau_c}
\frac{f(g)^2}{g^\delta (t/\tau_c)^{1+\gamma}}
\ \ \ \ (g\rightarrow +\infty\ {\rm then}\ t\rightarrow +\infty),
\end{equation}
where we have used the reflection formula $\Gamma(-z)\Gamma(z+1)=-\pi/\sin(\pi z)$.

On the other hand, Eq.\ (\ref{eq3.8}) with $\mu =2$ and $I_2(g)$ given by Eq.\ (\ref{eq5.16}) lead to
\begin{equation}\label{eq5.23}
p(g,t)\sim\frac{\sin(\pi\gamma /2)\Gamma(1+\gamma /2)}{\pi ac^2\tau_c}
\frac{f(g)^2}{g^{2\delta} (t/\tau_c)^{1+\gamma /2}}
\ \ \ \ (t\rightarrow +\infty\ {\rm then}\ g\rightarrow +\infty).
\end{equation}
Equations\ (\ref{eq5.22}) and\ (\ref{eq5.23}) suggest the scaling form
\begin{equation}\label{eq5.24}
p(g,t)\sim\frac{af(g)^2}{\tau_c (acg^\delta )^{3+2/\gamma}}
K\left(\frac{(t/\tau_c)^{\gamma/2}}{acg^\delta}\right)
\ \ \ \ (g\rightarrow +\infty\ {\rm and}\ t\rightarrow +\infty),
\end{equation}
with
\begin{equation}\label{eq5.25}
K(y)\sim\left\lbrace
\begin{array}{ll}
\mathcal{D}_{\rm I} y^{-2(1+1/\gamma)},&(y\rightarrow 0) \\
\mathcal{D}_{\rm II} y^{-1-2/\gamma},&(y\rightarrow +\infty)
\end{array}\right.
\end{equation}
where
\begin{equation}\label{eq5.26}
\mathcal{D}_{\rm I}=\sin(\pi\gamma)\frac{\Gamma(1+\gamma)}{\pi},
\end{equation}
and
\begin{equation}\label{eq5.27}
\mathcal{D}_{\rm II}=\sin(\pi\gamma/2)\frac{\Gamma(1+\gamma/2)}{\pi}.
\end{equation}
Note that $(t/\tau_c)^{\gamma/2}\sim acg^\delta$ corresponds to the cross-over from a `concentration' -- or `one-step' --  regime (for $(t/\tau_c)^{\gamma/2}< acg^\delta$) where the walker get stuck for a long time $t$ at the second maximum and then jumps directly to the first maximum, to a `many-steps' regime (for $(t/\tau_c)^{\gamma/2}> acg^\delta$) where she/he travels a long walk of total duration $t$ (with many steps) between the second and the first maxima. It remains to check whether or not the scaling form\ (\ref{eq5.24}) really exists.
%
%%%%%%%%%%%%%%%%%%%%
%
\subsubsection{Uniform expression of $p(g,t)$ for large $g$ and $t$}\label{sec5.3.1}
Rewriting\ (\ref{eq2.7}) as
\begin{equation}\label{eq5.28}
\phi(\lambda ,s)=\phi(\lambda ,1)\exp\left(-\frac{\lambda}{\pi}
\int_0^{+\infty}\frac{\ln\lbrack 1+(1-s)\hat{F}(k)\rbrack}{k^2+\lambda^2}\, dk\right) ,
\end{equation}
in the first line of\ (\ref{eq5.15}), with $\hat{F}(k)=\hat{f}(k)/(1-\hat{f}(k))$, one finds after some straightforward algebra,
\begin{eqnarray}\label{eq5.29}
\tilde{p}(g,\hat{\Psi}(q))&\sim&
\frac{f(g)^2}{cg^\delta}\, \left\lbrack\hat{\Psi}(q)-\frac{(\tau_c q)^{\gamma/2}}{\pi acg^\delta}
\int_0^{+\infty}\ln\left(1+\frac{1}{\overline{k}^2}\right)\, d\overline{k}\right\rbrack \nonumber \\
&=&\frac{f(g)^2}{cg^\delta}\, \left(\hat{\Psi}(q)-\frac{(\tau_c q)^{\gamma/2}}{acg^\delta}\right)
\ \ \ \ (g\rightarrow +\infty\ {\rm and}\ q\rightarrow 0),
\end{eqnarray}
where we have made the change of variable $k=(\tau_c q)^{\gamma/2}\overline{k}/a$ in the integral over $k$ in Eq.\ (\ref{eq5.28}) and use the fact that the integral over $\overline{k}$ is equal to $\pi$. Inverse Laplace transforming\ (\ref{eq5.29}) w.r.t. $q$ and using\ (\ref{eq5.27}) yields
\begin{equation}\label{eq5.30}
p(g,t)\sim\frac{f(g)^2}{cg^\delta}\, \left(\Psi(t)+
\frac{\mathcal{D}_{\rm II}}{\tau_c (acg^\delta )(t/\tau_c)^{1+\gamma/2}}\right)
\ \ \ \ (g\rightarrow +\infty\ {\rm and}\ t\rightarrow +\infty).
\end{equation}
Equation\ (\ref{eq5.30}) gives a uniform expression of $p(g,t)$ for a fast decreasing $f(\eta)$ when both $g$ and $t$ are large.

For $\gamma <1$, one has the large $t$ behavior $\Psi(t)\sim \mathcal{D}_{\rm I}\tau_c^{-1}(t/\tau_c)^{-1-\gamma}$ from which it is readily seen that\ (\ref{eq5.30}) reduces to the scaling form\ (\ref{eq5.24}) with the scaling function
\begin{equation}\label{eq5.31}
K(y)=\frac{1}{y^{1+2/\gamma}}\, \left(\frac{\mathcal{D}_{\rm I}}{y}+\mathcal{D}_{\rm II}\right) ,
\end{equation}
which fulfills the large and small argument behaviors\ (\ref{eq5.25}), as it should be.

For $\gamma =1$, there is no scaling form such as\ (\ref{eq5.24}) but the uniform expression\ (\ref{eq5.30}) makes it possible to determine the domains in the $(t,g)$ plane corresponding to the `concentration' -- or `one-step' --  regime and to the `many-steps' regime, respectively. Taking for instance $\Psi(t)=\tau_c^{-1}\exp(-t/\tau_c)$ and comparing the two terms on the right-hand side of\ (\ref{eq5.30}), one finds that the `one-step' regime corresponds to the domain
\begin{equation}\label{eq5.32}
acg^\delta >\frac{\exp(t/\tau_c)}{2\sqrt{\pi}\, (t/\tau_c)^{3/2}},
\end{equation}
and the `many-steps' regime to the complementary domain (i.e. with $>$ replaced with $<$).

\begin{figure}[htbp]
\begin{center}
\includegraphics [width=9cm] {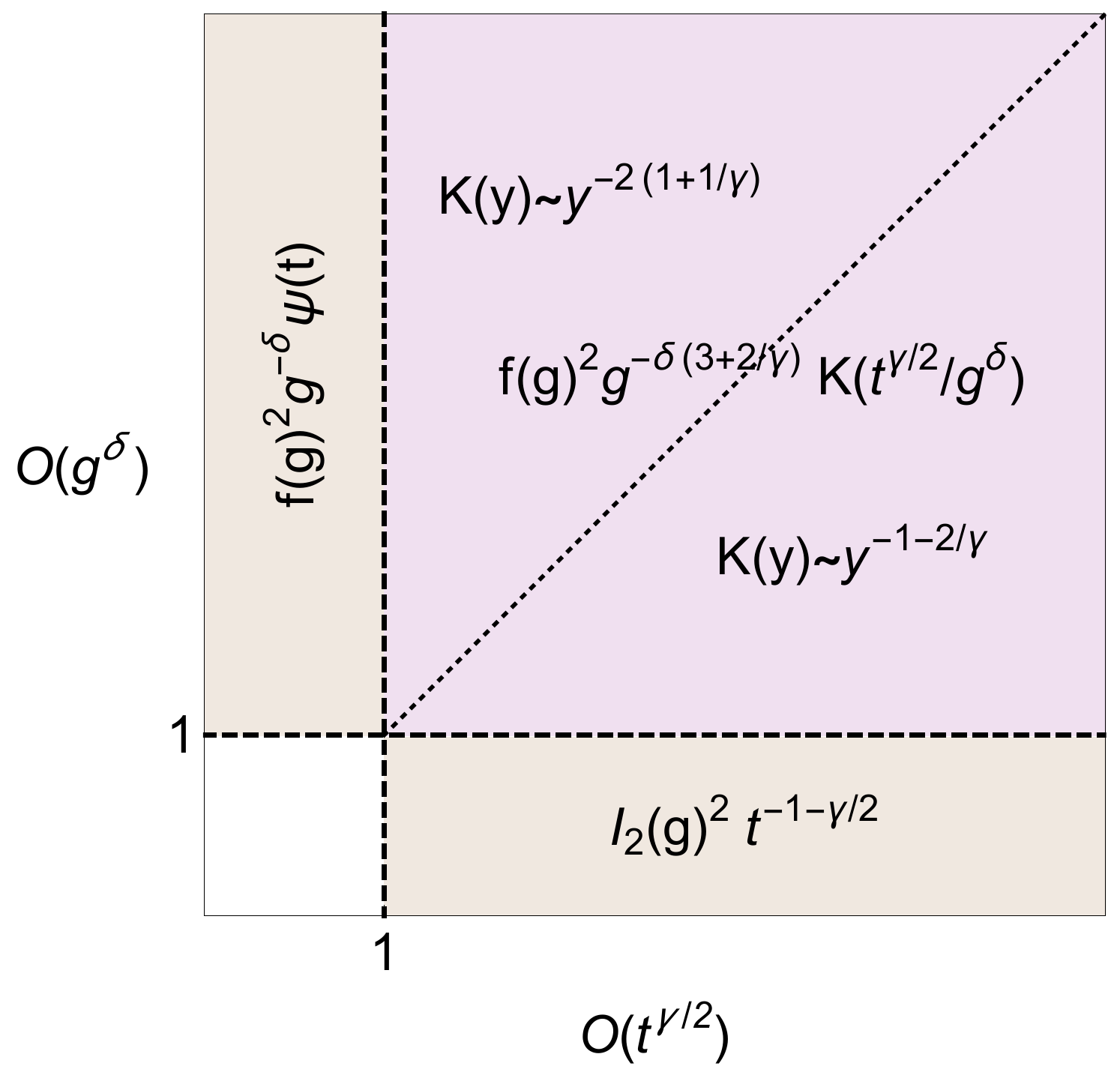}
\caption{\textsl{\it Schematic representation of the asymptotic behaviors of $p(g,t)$ for fast decreasing jump distributions and $0<\gamma <1$ (amplitudes are given in the text). For $t\gg 1$ and $g\lesssim O(1)$, $p(g,t)$ is given by Eq.\ (\ref{eq3.8}) with $\mu =2$. For both $t\gg 1$ and $g\gg 1$, one has the scaling form\ (\ref{eq5.24}) with $K(y)$ respectively given by the first line of\ (\ref{eq5.25}) if $t^{\gamma/2}\ll g^\delta$ and by the second line of\ (\ref{eq5.25}) if $t^{\gamma/2}\gg g^\delta$. For $t\lesssim O(1)$ and $g\gg 1$, $p(g,t)$ is given by Eq.\ (\ref{eq5.21}).}}
\label{fig1}
\end{center}
\end{figure}

Figure\ \ref{fig1} summarizes the different behaviors of $p(g,t)$ in the plane $(t^{\gamma/2},g^\delta )$ for a fast decreasing jump distribution and $0<\gamma <1$. The `one-step' (resp. `many-steps') regime is on the left (resp. right) of the diagonal.
\bigskip
%
%
%%%%%%%%%%%%%%%%%%%%
%
\section{Scaling form of the asymptotic behavior of $\bm{p(g,t)}$ for a L\'evy flight at large $\bm{t}$ and $\bm{g}$}\label{sec6}
In the preceding section we have found that for a fast decreasing jump distribution and $0<\gamma <1$, the asymptotic behavior of $p(g,t)$ at large $t$ and $g$ takes the scaling form\ (\ref{eq5.24}). Here we consider a L\'evy flight of index $0<\mu <2$ corresponding to a slow decreasing jump distribution. We will see that, in this case too, $p(g,t)$ at large $t$ and $g$ can be put into a scaling form for all $0<\gamma\le 1$.

Consider first the case $\gamma =1$. By Eq.\ (\ref{eq3.3}) and the scaling form (87) in\ \cite{MMS2014}, $p(g,t)$ depends on $t$ through the combination $tg^{-\mu}$ only as both $t$ and $g$ are large. Now, for $0<\gamma <1$, the comparison of Eqs.\ (\ref{eq3.2}) and\ (\ref{eq3.5}) indicates that $t$ should be replaced with $t^\gamma$. Thus,  for $0<\gamma\le 1$, $p(g,t)$ is expected to depend on  $t$ through the combination $t^\gamma g^{-\mu}$ only as both $t$ and $g$ are large. The simplest scaling ansatz for $p(g,t)$ consistent with this argument and the asymptotic behaviors\ (\ref{eq4.11}) and\ (\ref{eq5.6a}) reads
\begin{equation}\label{eq6.1}
p(g,t)\sim\frac{1}{a\tau_c}\, \left(\frac{a}{g}\right)^{1+\mu (1+1/\gamma)}
F_{\mu,\, \gamma}\left(\frac{a^\mu t^\gamma}{g^\mu \tau_c^\gamma}\right)
\ \ \ \ (g\rightarrow +\infty\ {\rm and}\ t\rightarrow +\infty),
\end{equation}
with the following large and small argument behaviors for the scaling function:
\begin{equation}\label{eq6.2}
F_{\mu,\, \gamma}(y)\sim\left\lbrace
\begin{array}{ll}
\mathcal{C}_{\rm I}\, y^{-1/\mu -1/\gamma},&1-{\rm Int}(\gamma) <\mu<2, \\
\mathcal{C}_{\rm II}\, y^{-1 -1/\gamma}\ln y,&\mu =1, \\
\mathcal{C}_{\rm III}\, y^{-1 -1/\gamma},&0<\mu<1,
\end{array}\right.
\ \ \ \ (y\rightarrow +\infty),
\end{equation}
and
\begin{equation}\label{eq6.3}
F_{\mu,\, \gamma}(y)\sim\mathcal{C}_{\rm IV}\, y^{1/2-1/\gamma},
\ \ \ \ (y\rightarrow 0).
\end{equation}
The term ${\rm Int}(\gamma)$ in the first line of\ (\ref{eq6.2}) denotes the integer part of $\gamma$ and the last two lines are for $0<\gamma <1$ only. The scaling form\ (\ref{eq6.1}) holds in the regime where both $g$ and $t$ are large, with $t^\gamma g^{-\mu}$ fixed.

\begin{figure}[htbp]
\begin{center}
\includegraphics [width=9cm] {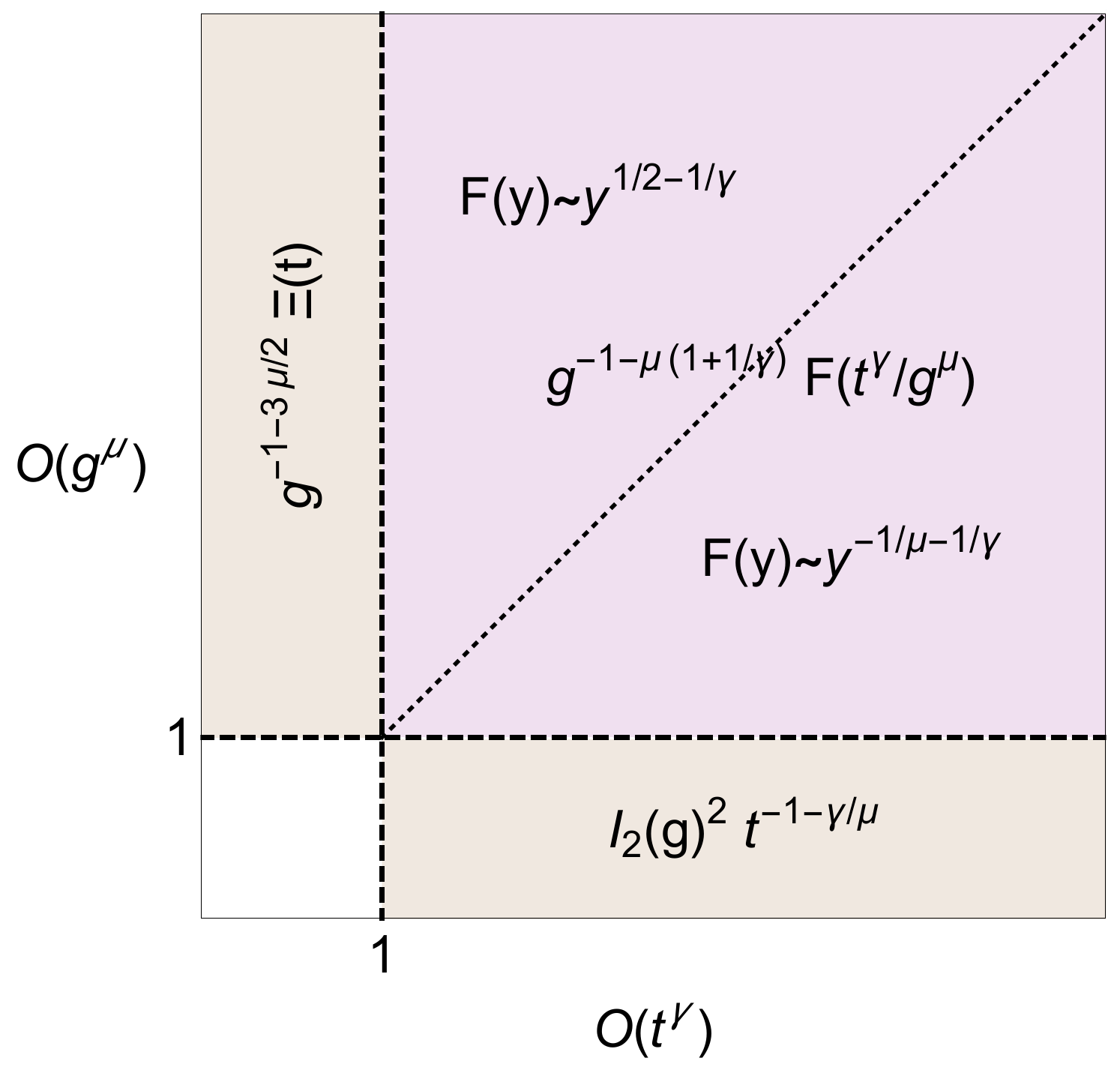}
\caption{\textsl{\it Schematic representation of the asymptotic behaviors of $p(g,t)$ for a L\'evy flight of index $1-{\rm Int}(\gamma) <\mu<2$ and $0<\gamma\le 1$ (amplitudes are given in the text). For $t\gg 1$ and $g\lesssim O(1)$, $p(g,t)$ is given by the first line of Eq.\ (\ref{eq3.17}). For both $t\gg 1$ and $g\gg 1$, one has the scaling form\ (\ref{eq6.1}) with $F(y)$ given by the first line of\ (\ref{eq6.2}) if $t^\gamma\gg g^\mu$ and by\ (\ref{eq6.3}) if $t^\gamma\ll g^\mu$. For $t\lesssim O(1)$ and $g\gg 1$, $p(g,t)$ is given by Eq.\ (\ref{eq5.4a}).}}
\label{fig2a}
\end{center}
\end{figure}

\begin{figure}[htbp]
\begin{center}
\includegraphics [width=9cm] {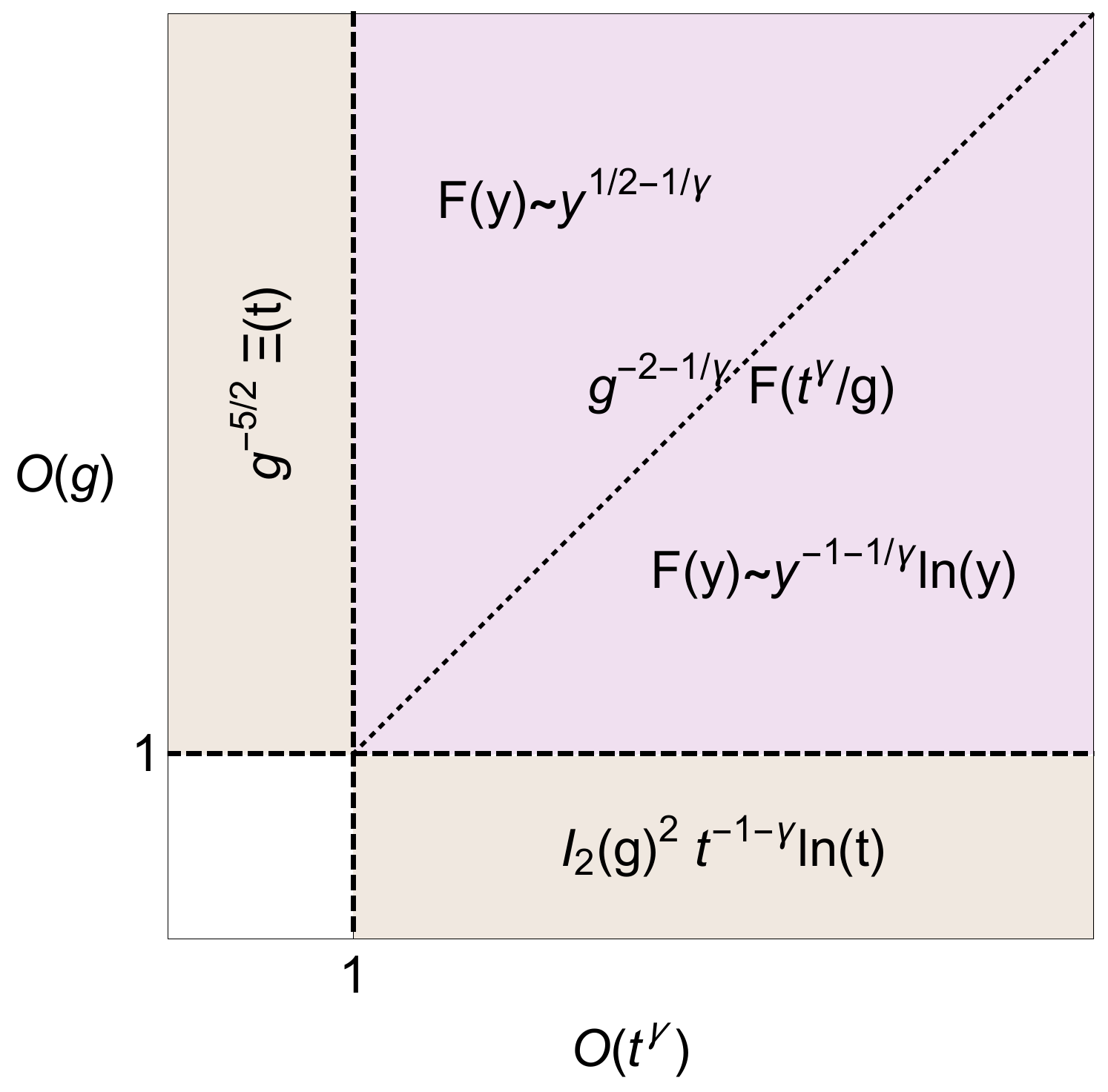}
\caption{\textsl{\it Schematic representation of the asymptotic behaviors of $p(g,t)$ for a L\'evy flight of index $\mu =1$ and $0<\gamma <1$ (amplitudes are given in the text). For $t\gg 1$ and $g\lesssim O(1)$, $p(g,t)$ is given by the second line of Eq.\ (\ref{eq3.17}). For both $t\gg 1$ and $g\gg 1$, one has the scaling form\ (\ref{eq6.1}) with $F(y)$ given by the second line of\ (\ref{eq6.2}) if $t^\gamma\gg g$ and by\ (\ref{eq6.3}) if $t^\gamma\ll g$. For $t\lesssim O(1)$ and $g\gg 1$, $p(g,t)$ is given by Eq.\ (\ref{eq5.4a}) with $\mu =1$.}}
\label{fig2b}
\end{center}
\end{figure}

\begin{figure}[htbp]
\begin{center}
\includegraphics [width=9cm] {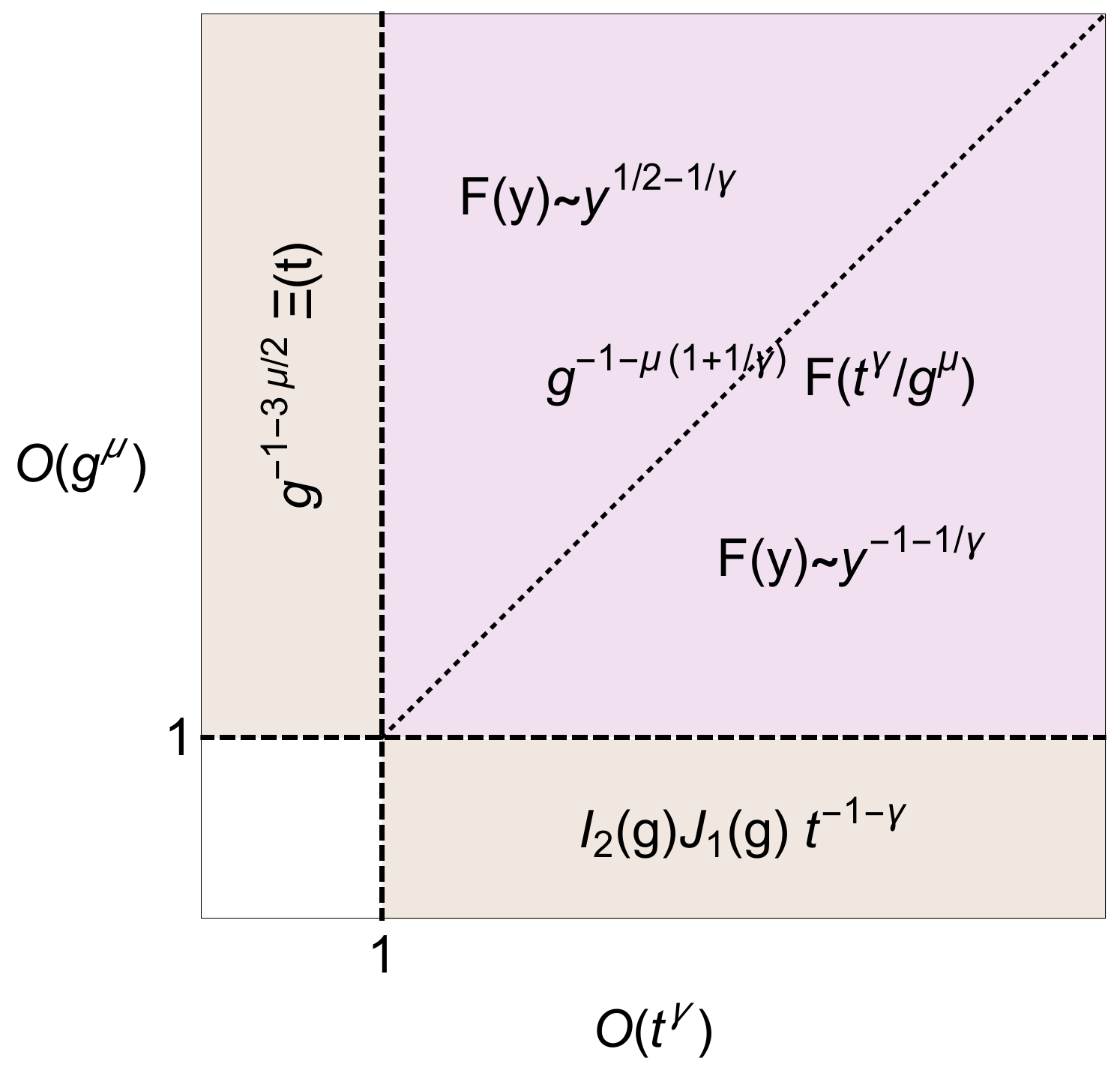}
\caption{\textsl{\it Schematic representation of the asymptotic behaviors of $p(g,t)$ for a L\'evy flight of index $0<\mu<1$ and $0<\gamma <1$ (amplitudes are given in the text). For $t\gg 1$ and $g\lesssim O(1)$, $p(g,t)$ is given by the third line of Eq.\ (\ref{eq3.17}). For both $t\gg 1$ and $g\gg 1$, one has the scaling form\ (\ref{eq6.1}) with $F(y)$ given by the third line of\ (\ref{eq6.2}) if $t^\gamma\gg g^\mu$ and by\ (\ref{eq6.3}) if $t^\gamma\ll g^\mu$. For $t\lesssim O(1)$ and $g\gg 1$, $p(g,t)$ is given by Eq.\ (\ref{eq5.4a}).}}
\label{fig2c}
\end{center}
\end{figure}

Figures\ \ref{fig2a},\ \ref{fig2b}, and\ \ref{fig2c} summarize the different behaviors of $p(g,t)$ in the plane $(t^\gamma,g^\mu )$ for a a L\'evy flight of index $0<\mu <2$. They correspond respectively to the cases $1-{\rm Int}(\gamma) <\mu<2$ with $0<\gamma\le 1$, $\mu =1$ with $0<\gamma <1$ and $0<\mu <1$ with $0<\gamma <1$.
\bigskip
\bigskip
\bigskip
%
%
%%%%%%%%%%%%%%%%%%%%
%
\section{Large $\bm{t}$ behavior of the marginal distribution $\bm{p_{\rm time}(t)}$}\label{sec7}
We can now determine the large $t$ behavior of the marginal distribution $p_{\rm time}(t)$ of the time between the first two maxima. The marginal distribution $p_{\rm gap}(g)$ of the gap between the first two maxima does not depend on $\gamma$ and is the same as the one already studied in Sec. 3 of Ref.\ \cite{MMS2014} in the discrete time RW setting.
\subsection{$\bm{\gamma =1}$}\label{sec7.1}
For $\gamma =1$, the large $t$ behavior of $p_{\rm time}(t)$ is readily obtained by integrating\ (\ref{eq3.3}) over $g$ and by using the large $l$ asymptotics (112) in\ \cite{MMS2014}. One finds
\begin{equation}\label{eq7.1}
p_{\rm time}(t)\sim\frac{1}{\tau_c}\times\left\lbrace
\begin{array}{ll}
\mathcal{A}_{\rm I}\, (t/\tau_c)^{-1-1/\mu},&1<\mu\le 2, \\
\mathcal{A}_{\rm II}\, \ln(t/\tau_c)\, (t/\tau_c)^{-2},&\mu=1, \\
\mathcal{A}_{\rm III}\, (t/\tau_c)^{-2},&0<\mu<1,
\end{array}\right.
\ \ \ \ (t\rightarrow +\infty),
\end{equation}
with
\begin{equation}\label{eq7.1a}
\begin{array}{l}
\mathcal{A}_{\rm I}=(\pi a)^{-1}\Gamma(1+1/\mu)\, 
\int_0^{+\infty}I_2(g)^2 dg, \\
\mathcal{A}_{\rm II}=\pi^{-2}, \\
\mathcal{A}_{\rm III}=\mu^{-1}\int_0^{+\infty}yF_{\mu,\, 1}(y)\, dy .
\end{array}
\end{equation}
%
%%%%%%%%%%%%%%%%%%%%
%
\subsection{$\bm{0<\gamma <1}$}\label{sec7.2}
To determine the large $t$ behavior of $p_{\rm time}(t)$ for $0<\gamma <1$, fix $\Lambda_1\gg 1$, $\Lambda_2=O(1)$, and for $t>(\Lambda_1/\Lambda_2)^{\mu/\gamma}$ write
\begin{equation}\label{eq7.2}
p_{\rm time}(t)=\int_0^{+\infty}p(g,t)\, dg=\int_0^{\Lambda_1}p(g,t)\, dg+
\int_{\Lambda_1}^{\Lambda_2 t^{\gamma/\mu}}p(g,t)\, dg+
\int_{\Lambda_2 t^{\gamma/\mu}}^{+\infty}p(g,t)\, dg.
\end{equation}

For $1<\mu\le 2$ and $t\rightarrow +\infty$,\ (\ref{eq7.2}) is dominated by the first two integrals in which one can use the large $t$ expression\ (\ref{eq3.8}) of $p(g,t)$. One finds
\begin{eqnarray}\label{eq7.3}
p_{\rm time}(t)&\sim&\frac{\sin(\pi\gamma /\mu)}{\sin(\pi /\mu)}\, 
\frac{\Gamma(1+\gamma /\mu)}{\pi a\tau_c\, (t/\tau_c)^{1+\gamma /\mu}}
\, \int_0^{\Lambda_2 t^{\gamma/\mu}}I_2(g)^2 dg \nonumber \\
&\sim&\frac{\sin(\pi\gamma /\mu)}{\sin(\pi /\mu)}\, 
\frac{\Gamma(1+\gamma /\mu)}{\pi a\tau_c\, (t/\tau_c)^{1+\gamma /\mu}}
\, \int_0^{+\infty}I_2(g)^2 dg\ \ \ \ (t\rightarrow +\infty).
\end{eqnarray}

For $0<\mu <1$ and $t\rightarrow +\infty$,\ (\ref{eq7.2}) is dominated by the contribution of large, $O(t^{\gamma/\mu})$, values of $g$, i.e. by the last two integrals in which one can use the large $t$ and $g$ scaling form\ (\ref{eq6.1}) of $p(g,t)$. One gets
\begin{eqnarray}\label{eq7.4}
p_{\rm time}(t)&\sim&\frac{1}{a\tau_c}\, \int_{\Lambda_1}^{+\infty}
\left(\frac{a}{g}\right)^{1+\mu (1+1/\gamma)}
F_{\mu,\, \gamma}\left(\frac{a^\mu t^\gamma}{g^\mu \tau_c^\gamma}\right)\, dg \nonumber \\
&=&\frac{1}{\mu\tau_c}\, \frac{1}{(t/\tau_c)^{1+\gamma}}\, 
\int_0^{a^\mu t^\gamma/\Lambda_1^\mu \tau_c^\gamma} y^{1/\gamma}
F_{\mu,\, \gamma}(y)\, dg \nonumber \\
&\sim&\frac{\gamma\mathcal{C}_{\rm III}}{\mu\tau_c}\, 
\frac{\ln(t/\tau_c)}{(t/\tau_c)^{1+\gamma}}\ \ \ \ (t\rightarrow +\infty).
\end{eqnarray}
where we have made the change of variable $y=a^\mu t^\gamma/g^\mu \tau_c^\gamma$ and used the third line of\ (\ref{eq6.2}).

Finally, for $\mu =1$ and $t\rightarrow +\infty$ we use the second line of\ (\ref{eq6.2}) in the second line of\ (\ref{eq7.4}) with $\mu =1$, which gives
\begin{equation}\label{eq7.5}
p_{\rm time}(t)\sim\frac{\gamma^2\mathcal{C}_{\rm II}}{2\tau_c}\, 
\frac{\ln^2(t/\tau_c)}{(t/\tau_c)^{1+\gamma}}\ \ \ \ (t\rightarrow +\infty).
\end{equation}

To summarize, for $0<\gamma <1$ the large $t$ behavior of the marginal distribution $p_{\rm time}(t)$ is given by
\begin{equation}\label{eq7.6}
p_{\rm time}(t)\sim\frac{1}{\tau_c}\times\left\lbrace
\begin{array}{ll}
\mathcal{A}_{\rm I}^\prime\, (t/\tau_c)^{-1-\gamma/\mu},&1<\mu\le 2, \\
\mathcal{A}_{\rm II}^\prime\, \ln^2(t/\tau_c)\, (t/\tau_c)^{-1-\gamma},&\mu=1, \\
\mathcal{A}_{\rm III}^\prime\, \ln(t/\tau_c)\, (t/\tau_c)^{-1-\gamma},&0<\mu<1,
\end{array}\right.
\ \ \ \ (t\rightarrow +\infty),
\end{equation}
with
\begin{equation}\label{eq7.6a}
\begin{array}{l}
\mathcal{A}_{\rm I}^\prime=a^{-1}\mathcal{C}_{\rm I}\Gamma(1-\mu/2)^2
\int_0^{+\infty}I_2(g)^2 dg, \\
\mathcal{A}_{\rm II}^\prime=\gamma^2\mathcal{C}_{\rm II}/2, \\
\mathcal{A}_{\rm III}^\prime=\gamma\mathcal{C}_{\rm III}/\mu ,
\end{array}
\end{equation}
where $\mathcal{C}_{\rm I}=\mathcal{B}_{\rm I}\Gamma(1-\mu/2)^{-2}$, $\mathcal{C}_{\rm II}=\mathcal{B}_{\rm II}\pi^{-1}$, and $\mathcal{C}_{\rm III}=\mathcal{B}_{\rm II}\tan(\mu\pi/2)/2$, the amplitudes $\mathcal{B}_{\rm I}$ and $\mathcal{B}_{\rm II}$ being respectively given by\ (\ref{eq3.17a}) and\ (\ref{eq3.17b}).

It can be seen that the smaller $\gamma$ is, the slower $p_{\rm time}(t)$ decreases as $t$ increases. The third lines of Eqs.\ (\ref{eq7.1}) and\ (\ref{eq7.6}) reveal an interesting freezing, as a function of $\mu$, of the large $t$ behavior of $p_{\rm time}(t)$ as $\mu$ decreases past the critical value $\mu_c =1$. It is the continuous time counterpart of the freezing phenomenon pointed out in Sec. 5 of\ \cite{MMS2014} for discrete time random walks. In particular, it follows immediately from\ (\ref{eq7.1}) and\ (\ref{eq7.6}) that the first moment of $p_{\rm time}(t)$ is never defined. This means that although the typical size of $t$ is $O(1)$, its average always diverges with the total duration of the walk. Writing $T$ the total duration of the walk, Eqs.\ (\ref{eq7.1}) and\ (\ref{eq7.6}) yield the estimate
\begin{equation}\label{eq7.7}
\langle\vert t\vert\rangle\sim\left\lbrace
\begin{array}{ll}
T^{1-\gamma/\mu}&1<\mu\le 2 \\
T^{1-\gamma}(\ln T)^2&\mu =1 \\
T^{1-\gamma}\ln T&0<\mu <1
\end{array}\right.
 \ \ \ \ (T\rightarrow +\infty),
\end{equation}
for $0<\gamma\le 1$.
%
%%%%%%%%%%%%%%%%%%%%
%
\section{Numerical simulations}\label{sec8}
In this section, we check our main analytical results by comparing them with the ones of numerical simulations of the corresponding CTRW. In the simulations, the random jump variables $\eta_i$'s and the random time intervals between two successive jumps $\tau_i$'s were drawn from the following distributions. 

{\it Jump variables.} For $\mu=2$, the jumps were drawn either from a Gaussian distribution $f(\eta) = \exp(-\eta^2/2)/\sqrt{2\pi}$ -- when we investigate the joint PDF of $g$ and $t$ for fast decreasing jump distribution -- or 
from a symmetric exponential distribution $f(\eta) =(b/2) \exp{(-b|\eta|)}$ -- when we study the marginal distribution $p_{\rm time}(t)$ --
in which case the amplitudes ${\cal A}_{\rm I}$ and ${\cal A'}_{\rm I}$ in\ (\ref{eq1:result_p_of_t}) and\ (\ref{eq2:result_p_of_t}) can be evaluated exactly. On the other hand, for $0<\mu<2$, we chose for $f(\eta)$ a symmetric Pareto distribution of index $\mu$ of the form
\begin{eqnarray}\label{eq:Pareto_jump}
f(\eta) = \begin{cases}
&0 \;, \; {\rm for} \; |\eta| < 1 \; \\
&\dfrac{\mu}{2} \, |\eta|^{-1-\mu} \;, \; {\rm for} \; |\eta| \geq 1 \;,
\end{cases}
\end{eqnarray}
for which it is straightforward to compute the parameter $a$ in Eq. (\ref{def_mu}). One finds,
\begin{eqnarray}\label{def_a}
a = \left[ \Gamma(1-\mu) \cos{\left(\frac{\mu \pi}{2} \right)}\right]^{1/\mu} \;.
\end{eqnarray}
The reason for such a choice is twofold: (i) random numbers drawn from\ (\ref{eq:Pareto_jump}) are easy and fast to generate numerically (see for instance Ref. \cite{MRS2013}) and (ii) $f(\eta)$ is a power law for all $|\eta| \geq 1$ (instead of $|\eta| \gg 1$ ) which allows us to access the asymptotic regime characterizing L\'evy flights rather quickly in numerical simulations.  

{\it Time intervals.} Similarly, to generate a CTRW numerically, it is also convenient to draw the time variables $\tau_i$'s from a distribution $\Psi(\tau)$ which is a (one-sided) Pareto distribution of index $\gamma < 1$ of the form
 \begin{eqnarray}\label{eq:Pareto_time}
\Psi(\tau) = \begin{cases}
&0 \;, \; {\rm for} \; \tau < 1 \; \\
&\gamma \, \tau^{-1-\gamma} \;, \; {\rm for} \; \tau \geq 1 \;. 
\end{cases}
\end{eqnarray}
For such a distribution the characteristic time $\tau_c$ is given by
\begin{eqnarray}\label{tauc}
\tau_c = [\Gamma(1-\gamma)]^{1/\gamma} \;.
\end{eqnarray}
%
%%%%%%%%%%%%%%%%
%
\subsection{Scaling form of the asymptotic behavior of $\bm{p(g,t)}$ at large $\bm{t}$ and $\bm{g}$}\label{sec8.1}
Computing a joint PDF numerically is notoriously a quite hard task. So, instead of $p(g,t)$ we consider the cumulative distribution
\begin{eqnarray}\label{def_cumul}
p_>(g,t) = \int_t^\infty p(g,t') dt' \;,
\end{eqnarray}
which yields better statistics at a lesser cost.

We first present the result for a fast decreasing jump distribution $f(\eta)$ in the sense of Sec.\ \ref{sec5.3}, i.e. $f(\eta + x) \sim f(\eta) \exp(-c \, \eta^\delta \, x)$ when $\eta \to \infty$ [see Eq. (\ref{eq5.14})]. In this case, one expects from Eq. (\ref{scaling1a}) that in the limit $t \to \infty$ and $g \to \infty$ keeping $g \, t^{-\gamma/(2 \delta)}$ fixed, $p_>(g,t)$ takes the following scaling form:
\begin{eqnarray}\label{cumul_fast}
p_>(g,t) \sim a\, f(g)^2 \left(\frac{t}{\tau_c}\right)^{-3\gamma/2}  H\left(\frac{(a\,c)^{1/\delta} g}{(t/\tau_c)^{\gamma/(2 \delta)}} \right) \;, \;
\end{eqnarray}
where the scaling function $H(z)$ is related to the scaling function $K(z)$ through the relation
\begin{eqnarray}
H(z) = \frac{1}{z^{3 \delta}} \int_{1/z^{2\delta/\gamma}} K(y^{\gamma/2}) \, dy\;.
\end{eqnarray}
From the asymptotic behaviors of $K(z)$ in Eq. (\ref{scaling1b}), one deduces the asymptotic behaviors of $H(z)$ as follows
\begin{eqnarray}\label{asymptH}
H(z) \sim \begin{cases}
&\dfrac{2 \,D_{\rm II}}{\gamma} \; z^{-2 \delta} \;, \; (z \to 0) \\
&\dfrac{D_{\rm I}}{\gamma} \; z^{-\delta} \; \; (z \to \infty) \;.
\end{cases}
\end{eqnarray} 
As discussed above, the limit $z \to \infty$ corresponds to the `single-step' regime while the limit $z \to 0$ corresponds to the `many-steps' regime. 

\begin{figure}
\includegraphics[width=0.8\linewidth]{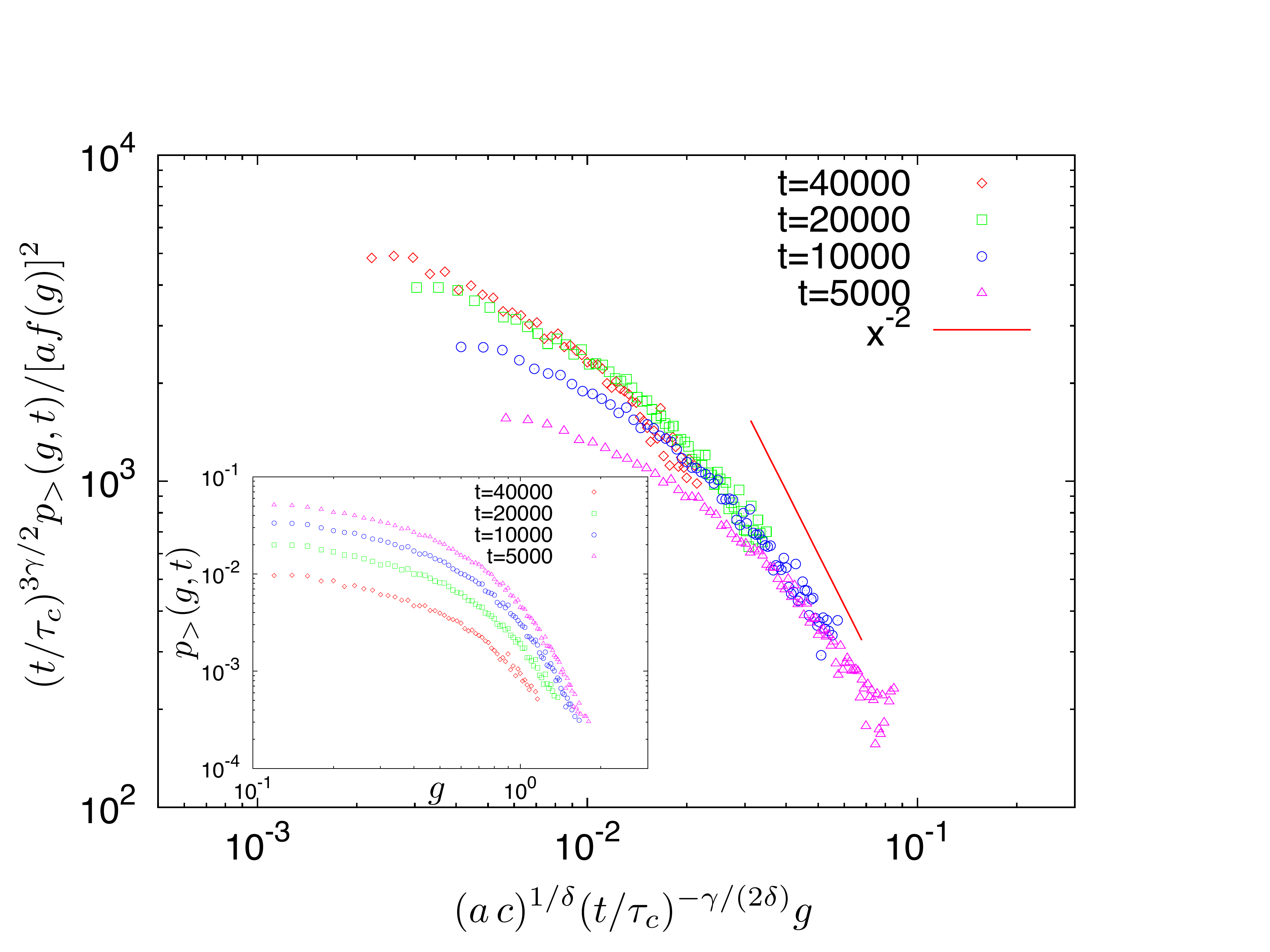}
\caption{Plot of $(t/\tau_c)^{3\gamma/2} p_>(g) [f(g)]^2$ as a function of $(a\, c)^{1/\delta} (t/\tau_c)^{-\gamma/(2\delta)} g$ for a fast decreasing jump distribution, namely a Gaussian jump distribution, corresponding to $\delta = 1$, and with $\gamma = 0.9$. The different curves correspond to different values of $t = 5000, 10000, 20000$ and $40000$ for a CTRW of duration $T = 10^5$. The (relatively) good collapse of the different curves for large values of $g$ corroborate the scaling form predicted in Eq. (\ref{cumul_fast}). The solid line is a guide to eyes, corresponding to the small argument behavior of the scaling function $H(z)$ in Eq. (\ref{asymptH}) with $\delta = 1$. {\bf Inset:} same data as in the main panel but without the rescaling.}\label{Fig:fast}
\end{figure}

In Fig. \ref{Fig:fast} we show a plot of $(t/\tau_c)^{3\gamma/2} p_>(g) [f(g)]^2$ as a function of $(a\, c)^{1/\delta} (t/\tau_c)^{-\gamma/(2\delta)} g$ for a fast decreasing jump distribution, namely a Gaussian jump distribution, corresponding to $\delta = 1$, and with $\gamma = 0.9$, for different values of $t = 5000, 10000, 20000$ and $40000$, while the total duration of the CTRW is $T = 10^5$. The data have been obtained by averaging over $10^6$ independent realizations of the CTRW. We see on this plot that, for sufficiently large values of $g$, the curves for different times $t$ tend to collapse on a master curve, as predicted by Eq. (\ref{cumul_fast}), which indeed holds only in the limit $g \to \infty$, $t \to \infty$ with $g \, t^{-\gamma/(2 \delta)}$ fixed. For smaller values of $g$ we see a clear deviation from this scaling form. We emphasize on the fact that, in our simulations, we have been able to access the regime where the scaling variable $\propto g \, t^{-\gamma/(2 \delta)}$ is small (corresponding to the `many-steps' regime), where $H(z)$ is expected to behave as $H(z) \propto z^{-2}$. It is extremely difficult to reach the complementary `single-step' regime corresponding to $g \, t^{-\gamma/(2 \delta)} \gg 1$. Indeed, for such high values of $t \geq 5000$ the condition $g \, t^{-\gamma/(2 \delta)} \gg 1$ with $\delta = 1$ and $\gamma = 0.9$ yields $g\gg 46$, which is already large enough to prevent any access to reliable statistics (to understand the difficulty, just realize that for the Gaussian jump distribution we consider in the `single-step' regime, the typical gaps are greater than $10$ less often than one time out of the Avogadro's number). Consequently, the clear numerical observation of the crossover behavior between the `many-steps' and the `single-step' regimes described by the scaling function in Eq. (\ref{cumul_fast}) goes beyond the scope of the present paper. Notice that, for $\gamma=1$ and $t\sim\tau_c$, the `single-step regime' was clearly demonstrated in Ref. \cite{MMS2014}.    

We now present numerical results for L\'evy flights ($0<\mu <2$), for which $f(\eta)$ is slow decreasing in the sense of Sec.\ \ref{sec5.1}. In this case, one expects from Eq. (\ref{scaling2a}) that in the limit $t \to \infty$ and $g \to \infty$ at fixed $g t^{-\gamma/\mu}$, $p_>(g,t)$ takes the following scaling form:
\begin{eqnarray}\label{scaling_cumul}
p_>(g,t) \sim \frac{1}{a} \left(\frac{t}{\tau_c} \right)^{-\gamma(1+1/\mu)} G_{\mu,\gamma}\left(\frac{g}{a} \left( \frac{\tau_c}{t}\right)^{\gamma/\mu} \right) \;,
\end{eqnarray}
where the function $G_{\mu,\gamma}(z)$ is related to the scaling function $F_{\mu,\gamma}(z)$ through the relation
\begin{eqnarray}
G_{\mu,\gamma}(z) = \frac{1}{z^{1+\mu}} \int_{1/z^{\mu/\gamma}}^\infty F_{\mu,\gamma}(y^\gamma) \, dy \;.
\end{eqnarray}
The small and large $z$ behaviors of $G_{\gamma,\mu}(z)$ are readily obtained from the asymptotic behaviors of $F_{\mu,\gamma}(z)$ given in Eqs. (\ref{scaling2b}) and (\ref{scaling2c}). One gets
\begin{eqnarray}\label{G:large}
G_{\mu,\gamma}(z) \sim \frac{{\cal C}'_{\rm III}}{z^{1+\mu}} \; \; (z \to \infty) \;,
\end{eqnarray}
for large $z$, where ${\cal C}'_{\rm III} = \int_0^\infty F_{\mu,\gamma}(y^\gamma) \, dy$, and
\begin{eqnarray}\label{G:small}
G_{\mu,\gamma}(z) \sim
\begin{cases}
&{{\cal C}_{\rm I}}\,{z^{-\mu}} \;, \; 1 -{\rm Int}(\gamma)< \mu < 2 \\
& \dfrac{{\cal C}_{\rm II}}{\gamma} z^{-1} \ln(1/z) \;, \; \mu =1 \\
& {{\cal C}_{\rm III}}\,{z^{-1}} \;, \; 0 < \mu < 1 \;.
\end{cases}\hfill \; (z \to 0)\;,
\end{eqnarray}
for small $z$, where the last two lines are for $0<\gamma <1$ only.

\begin{figure}[ht]
\includegraphics[width = 0.8\linewidth]{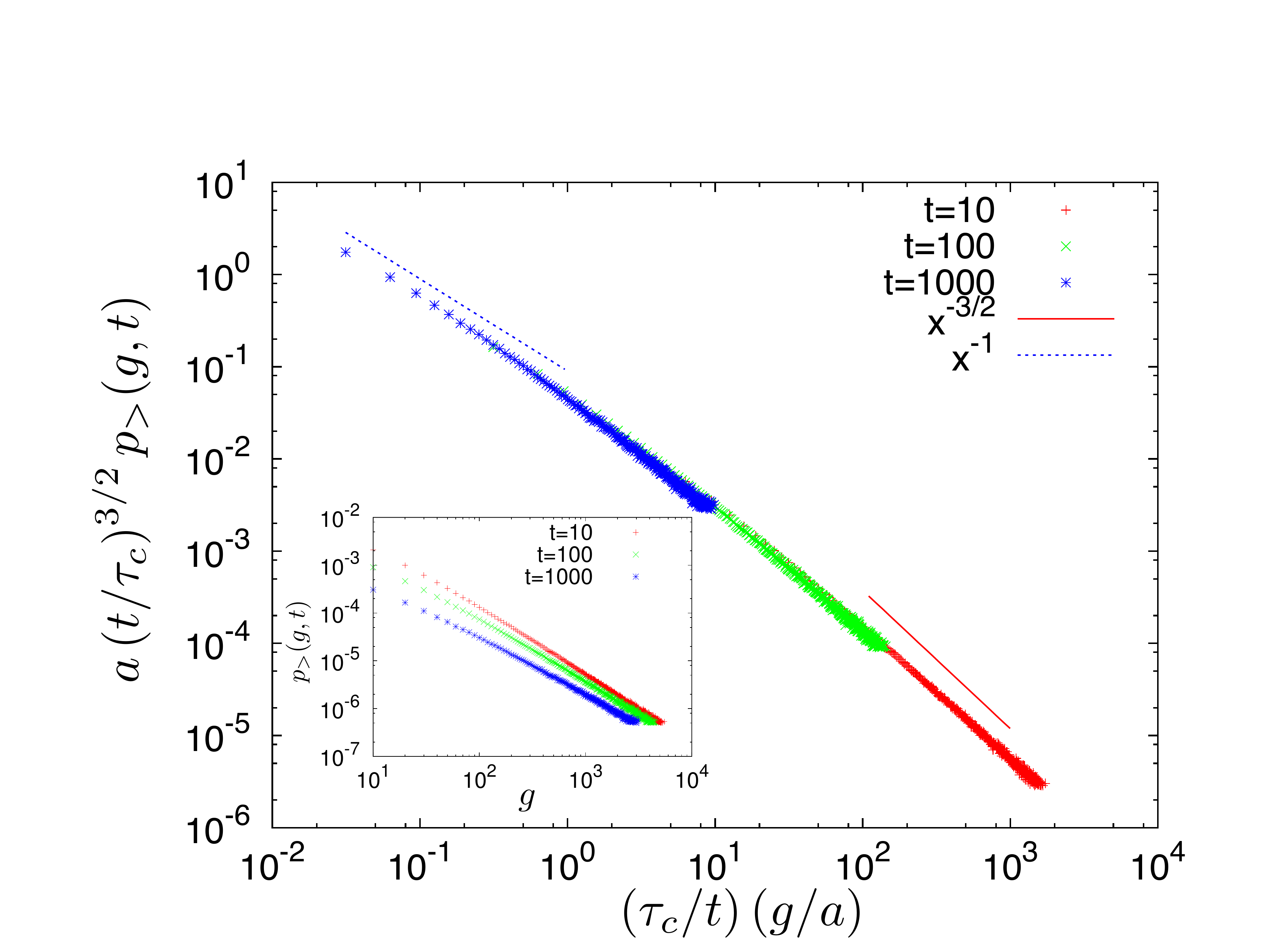}
\caption{Plot of $a\, (t/\tau_c)^{\gamma(1+1/\mu)}\, p_>(g,t)$ as a function of $(g/a) ({\tau_c}/{t})^{\gamma/\mu}$
with $\gamma = \mu = 1/2$ for different values of $t = 10, 100$ and $t=1000$, for a CTRW of total duration $T = 10^4$. The good collapse of the curves for different values of $t$ confirm the scaling form predicted in Eq. (\ref{scaling_cumul}). The dashed and solid lines are guides to the eyes, indicating the algebraic behaviors predicted in Eqs. (\ref{G:large}) and (\ref{G:small}). {\bf Inset:} same data as in the main panel but without the rescaling.}\label{fig_joint_combined}
\end{figure}
In Fig. \ref{fig_joint_combined} we show a scaled plot of $a\, (t/\tau_c)^{\gamma(1+1/\mu)}\, p_>(g,t)$ as a function of $(g/a) ({\tau_c}/{t})^{\gamma/\mu}$
with $\gamma = \mu = 1/2$ for different values of $t = 10, 100$ and $t=1000$, for CTRWs of total duration $T = 10^4$. The data have been obtained by averaging over $10^6$ independent realizations of the CTRW. This plot demonstrates a good agreement with our predictions for the scaling form of the joint PDF in Eq.~(\ref{scaling_cumul}) as well as with the asymptotic behaviors predicted in Eqs. (\ref{G:large}) and~(\ref{G:small}). 
%
%%%%%%%%%%%%%%%%
%
\subsection{Asymptotic behavior of $\bm{p_{\rm time}(t)}$ at large $\bm{t}$}\label{sec8.2}
An interesting feature of our results for the distribution $p_{\rm time}(t)$ of the time $t$ between the first two maxima is the presence of logarithmic corrections for $0<\mu \leq 1$ and $0<\gamma <1$ [see Eq. (\ref{eq2:result_p_of_t})]. To demonstrate these logarithmic corrections numerically, it is convenient to study the distribution $p_{\rm logtime}(Y)$ of the logarithm of the time, $Y = \ln  (t/\tau_c)$ instead of the time distribution itself. From $p_{\rm logtime}(Y) = \tau_c \e^{Y} p_{\rm time}(\tau_c \e^{Y})$ and Eq. (\ref{eq2:result_p_of_t}) it follows that the asymptotic behavior of $p_{\rm logtime}(Y)$ at large $Y$ and for $0<\gamma <1$ is given by
\begin{eqnarray}\label{eq:asympt_log}
p_{\rm logtime}(Y) \sim
\begin{cases}
&{\cal A}'_{\rm I} \, \exp{\left(-\dfrac{\gamma}{\mu}\, Y\right)} \;, \; 1 < \mu \leq 2 \\
&{\cal A}'_{\rm II} \, Y^2 \, \exp{\left(-\gamma \, Y\right)} \;, \, \mu = 1 \\
&{\cal A}'_{\rm III} \, Y  \exp{\left(-\gamma \, Y\right)} \;, 0 < \mu < 1 \;,
\end{cases}
\end{eqnarray}
with the amplitudes
\begin{eqnarray}
&&{\cal A}'_{\rm I} = \frac{1}{2 \pi} {\sin{\left( {\gamma \pi}/{2}\right)} \Gamma(1 + {\gamma}/{2})} \\
&&{\cal A}'_{\rm II} = \frac{\gamma^2}{2 \pi^3} \Gamma(1+\gamma)\sin{(\gamma \pi)} \;, \\
&&{\cal A}'_{\rm III} = \frac{\gamma}{2 \mu \pi^2} \tan{\left(\frac{\mu \pi}{2} \right)} \sin{(\gamma \pi)} \Gamma(1+\gamma) \;,
\end{eqnarray}
where the value of ${\cal A}'_{\rm I}$ corresponds to a symmetric exponential jump distribution (see the beginning of this section).

\begin{figure}[ht]
\includegraphics[width = 0.8\linewidth]{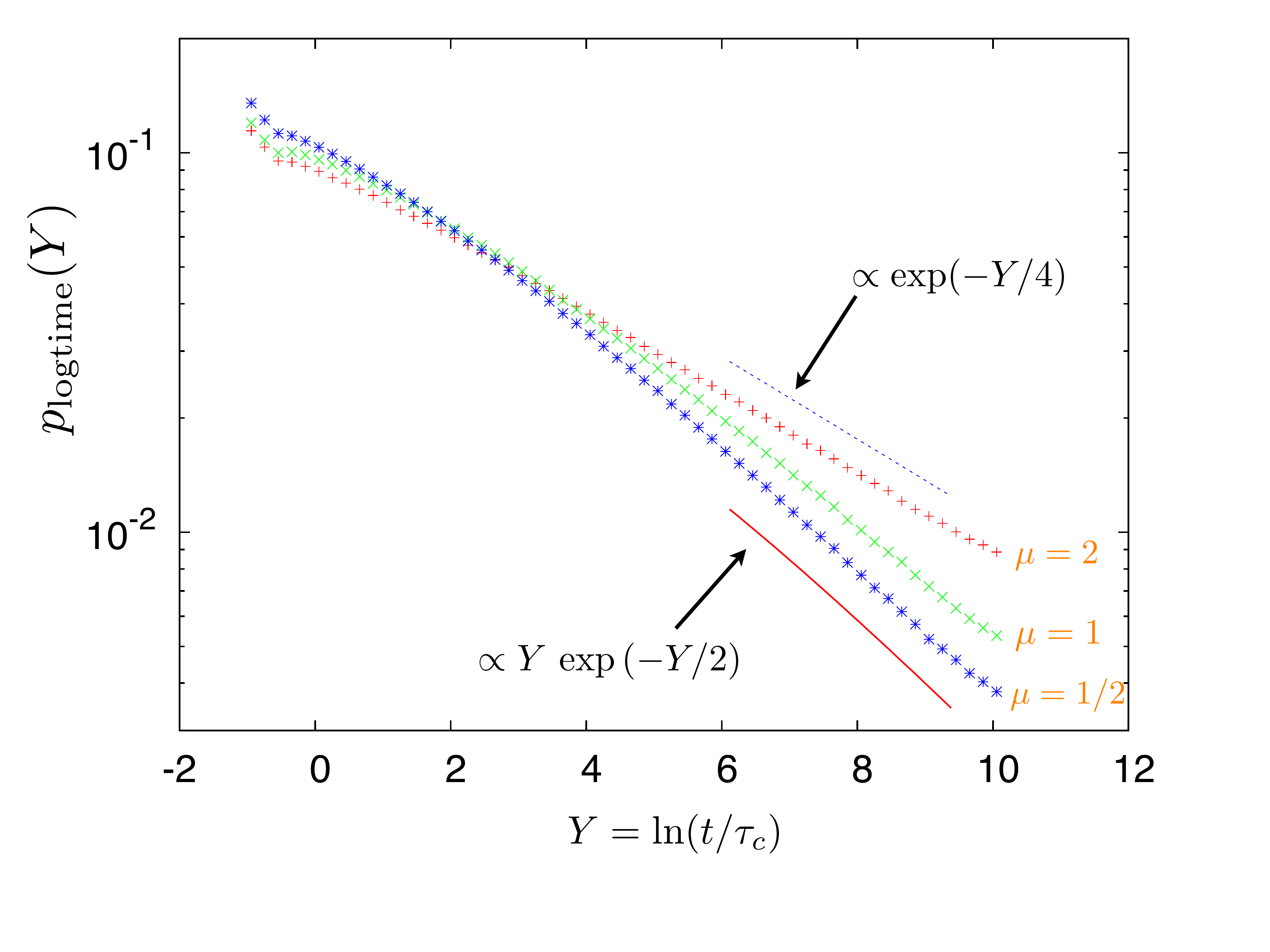}
\caption{Plot of $p_{\rm logtime}(Y)$ as a function of $Y = \ln (t/\tau_c)$ on a log-linear plot for $\gamma = 1/2$ and for different values of $\mu = 2, 1$ and $1/2$, corresponding to the three different regimes in Eq. (\ref{eq:asympt_log}). The solid lines are guides to the eyes corresponding to the asymptotic behaviors for $\mu = 2$ and $\mu = 1/2$ given in Eq. (\ref{eq:asympt_log}). These data have been obtained for CTRWs of length $T = 10^5$.}\label{fig_time1}
\end{figure}

\begin{figure}[ht]
\includegraphics[width = 0.8\linewidth]{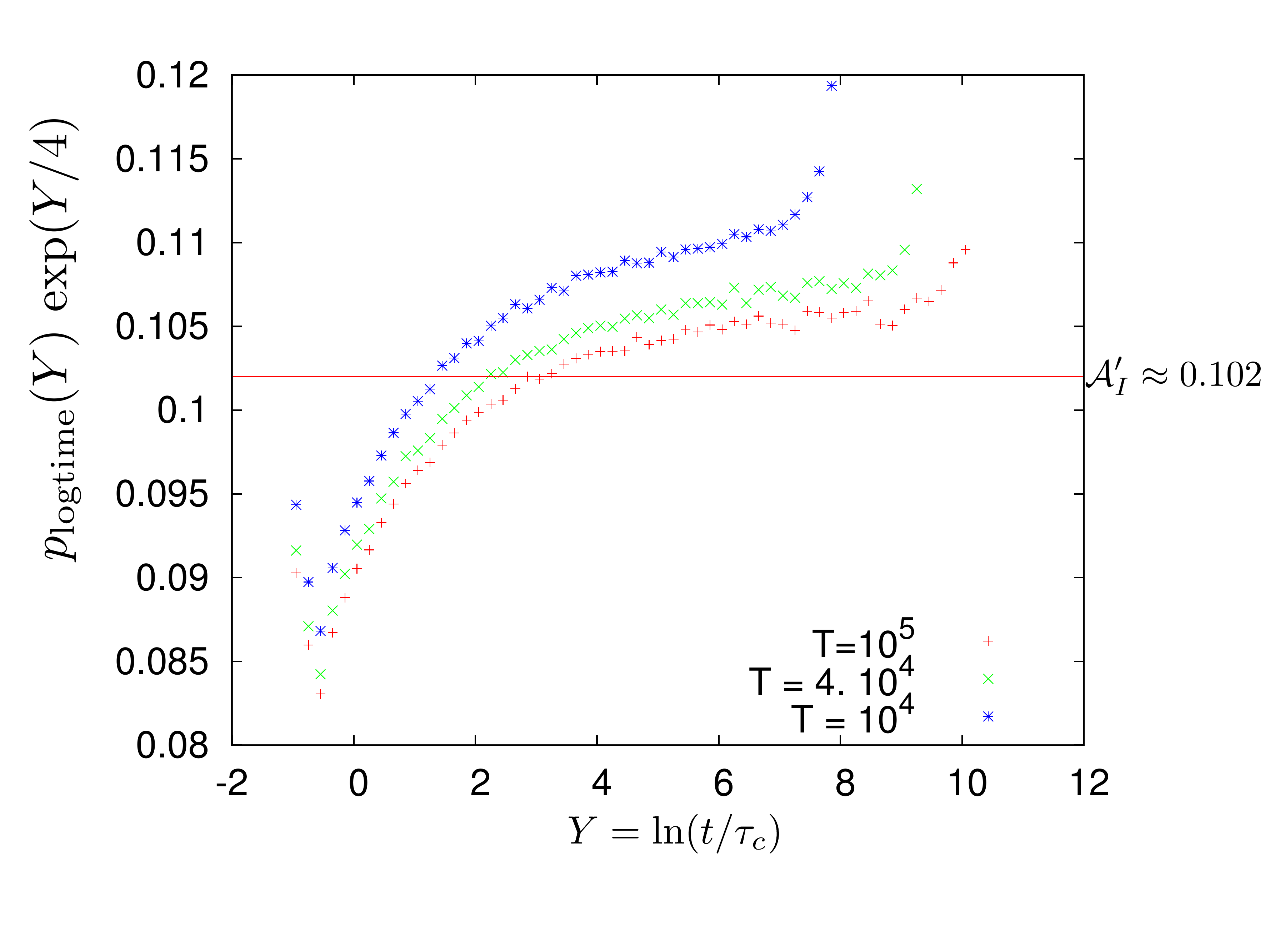}
\caption{Plot of $p_{\rm logtime}(Y)\, \exp(\gamma Y/\mu)$ for $\gamma = 1/2$ and $\mu =2$ (namely exponentially distributed jumps) as a function of $Y = \log(t/\tau_c)$ for CTRWs of increasing length, namely $T = 10^4$, $T=4\,\cdot 10^4$ and $T = 10^5$. Note that the upward bending is a finite size effect, as our predictions actually hold for $1 \ll t \ll T$ which is confirmed by the fact that this bending happens for larger and larger values of $t$ as the total duration $T$ of the CTRW increases. In the appropriate regime such that $1 \ll t \ll T$ $p_{\rm logtime}(Y)\, \exp(\gamma Y/\mu)$ converges to a constant which is indeed close (with an accuracy of less than 5$\%$ for $T = 10^5$) to our analytical predictions ${\cal A}'_{\rm I}$ in Eq. (\ref{eq:asympt_log}).}\label{fig_precise_exp}
\end{figure}

In Fig. \ref{fig_time1}, we show a plot of $p_{\rm logtime}(Y)$ as a function of $Y$ for three different values of $\mu$: $\mu=2$ (with exponentially distributed jumps), $\mu = 1$, and $\mu = 1/2$, that are representative of the three different regimes in Eq. (\ref{eq:asympt_log}), for a fixed value of $\gamma = 1/2$. This plot shows that the large $Y$ behavior of $p_{\rm logtime}(Y)$ does depend on $\mu$ as predicted in Eq. (\ref{eq:asympt_log}). To make a precise comparison between our numerical results and our analytical predictions (and in particular to demonstrate the expected logarithmic corrections), we now analyze these data, for different values of $\mu$, more carefully.

Fig. \ref{fig_precise_exp} corresponds to the case of a sub-diffusive walk with $\mu =2$ (with exponentially distributed jumps) and $\gamma =1/2$. We show a plot of $p_{\rm logtime}(Y)\, \exp(\gamma Y/\mu)$ for $\gamma = 1/2$ as a function of $Y = \log(t/\tau_c)$ for CTRWs of increasing length, namely $T = 10^4$, $T=4\, 10^4$ and $T = 10^5$. Here, the data have again been obtained by averaging over $10^6$ independent CTRWs. It can be seen that in the time interval $1 \ll t \ll T$ where our prediction actually holds, $p_{\rm logtime}(Y)\, \exp(\gamma Y/\mu)$ approaches a constant which gets closer to the expected value ${\cal A}'_{\rm I}$ as the total duration $T$ of the walk increases. The observed small upward discrepancy, as well as the steep upward bending for larger $Y$, are finite size effects. Note the very stretched vertical scale. For the longest walks with $T = 10^5$, numerical and analytical results are in agreement to within an accuracy of less than 5$\%$.

\begin{figure}[ht]
\includegraphics[width = 0.8\linewidth]{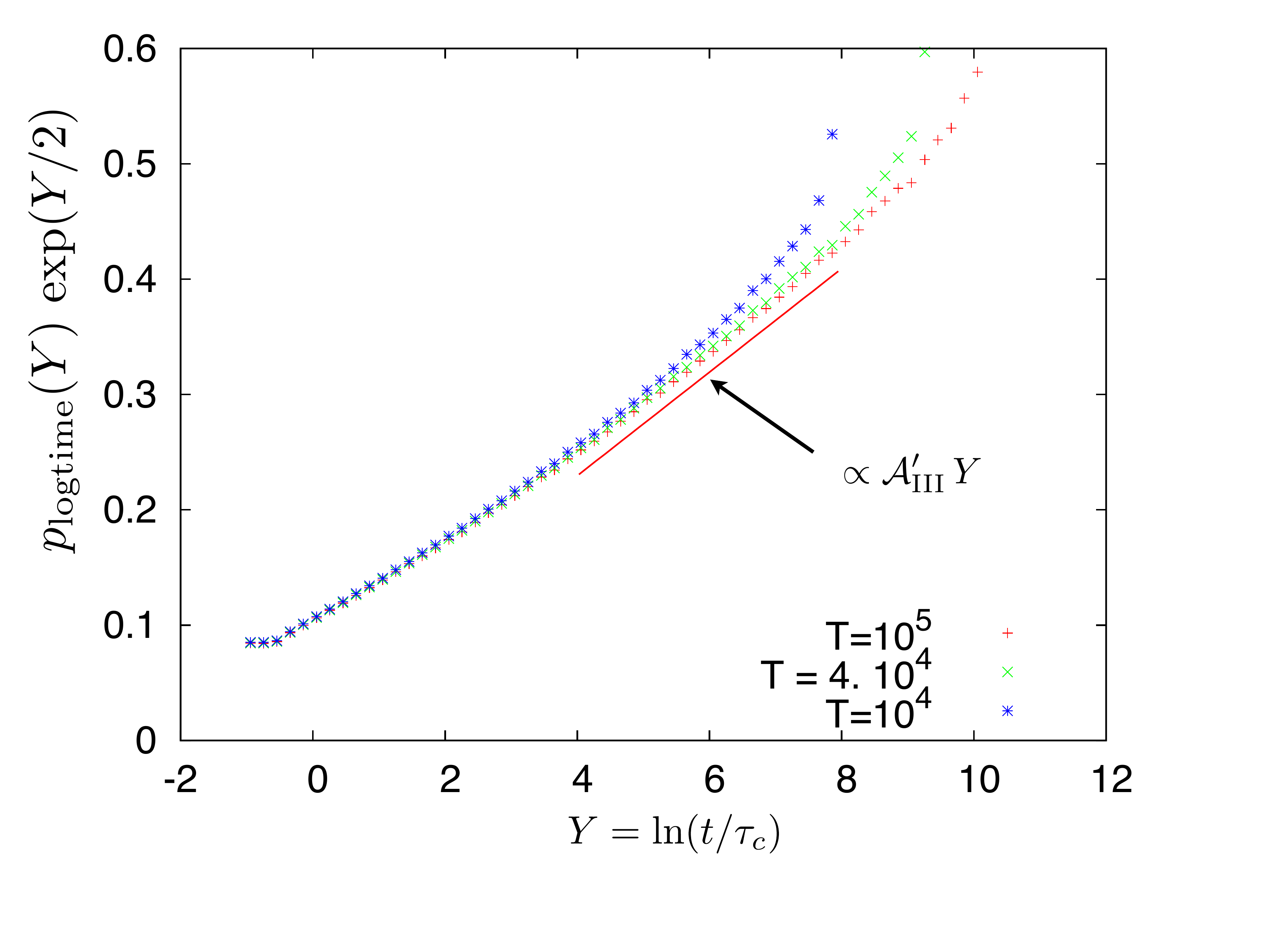}
\caption{Plot of $p_{\rm logtime}(Y) \exp{(\gamma Y)}$ as a function of $Y$ for $\mu = 1/2$ and $\gamma = 1/2$ and for different durations of the CTRW: $T = 10^4$, $T=4\,\cdot 10^4$, and $T = 10^5$. The solid line, which corresponds to our analytical result in Eq. (\ref{eq:asympt_log}) with ${\cal A}'_{\rm III} = 0.0448968 \ldots$ is a guide to eye, indicating a good agreement between our analytical results and numerics. This linear behavior is a clear signature of the logarithmic corrections in $p_{\rm time}(t)$, which is a peculiarity of the CTRW model. Upward bending at large $Y$ is a finite size effect which decreases as $T$ increases.}\label{fig_precise_III}
\end{figure} 

\begin{figure}[h]
\includegraphics[width = 0.8\linewidth]{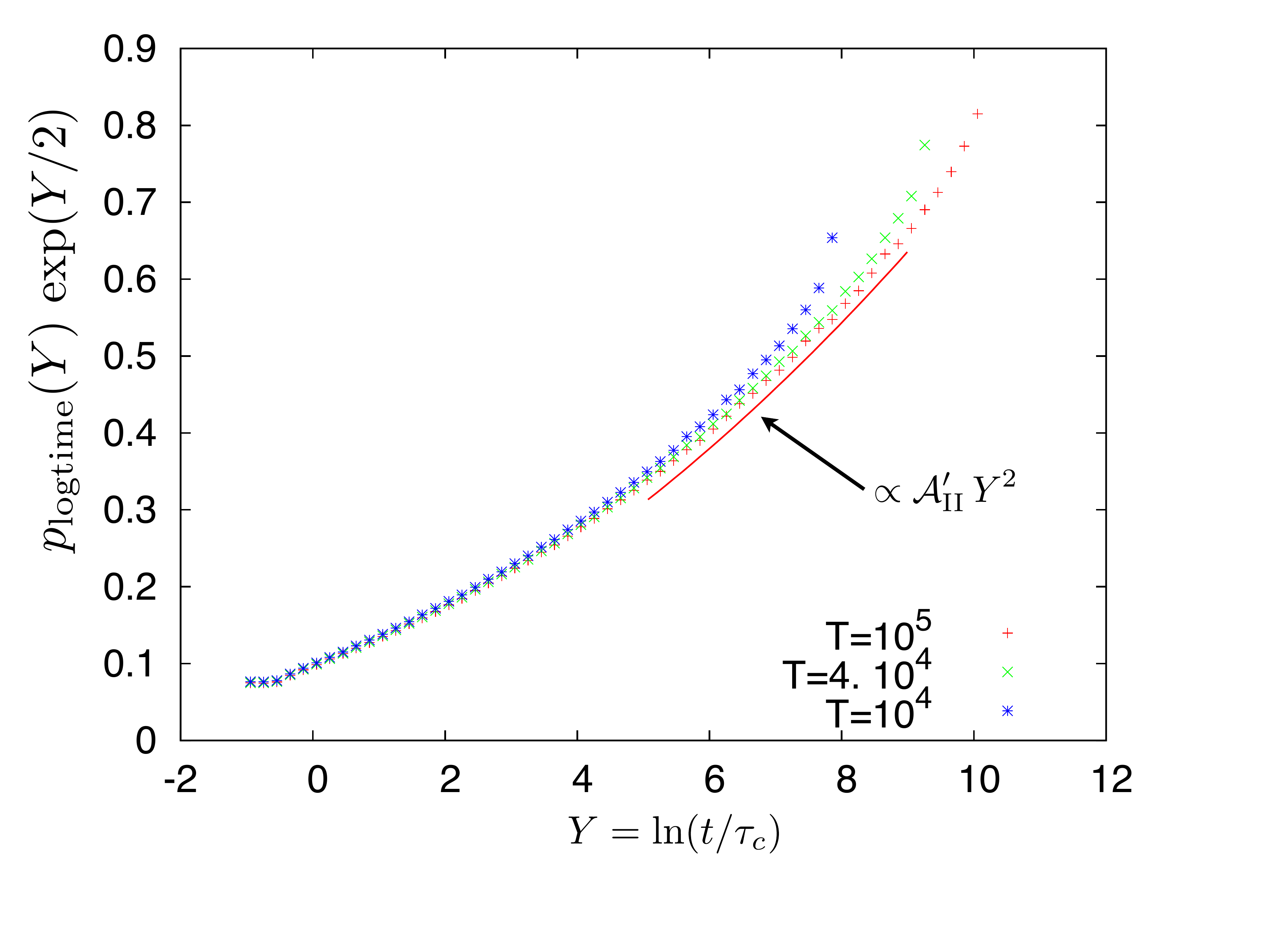}
\caption{Plot of $p_{\rm logtime}(Y) \exp{(\gamma Y)}$ as a function of $Y$ for $\mu = 1$ and $\gamma = 1/2$ and for different durations of the CTRW, $T = 10^4$, $T=4\, \cdot 10^4$, and $T = 10^5$. The solid line, which corresponds to our analytical result in Eq. (\ref{eq:asympt_log}) with ${\cal A}'_{\rm II} = 0.000357277 \ldots$ is a guide to eye, indicating a good agreement between our analytical results and numerics. Upward bending at large $Y$ is a finite size effect which decreases as $T$ increases.}\label{fig_precise_II}
\end{figure}

Next we consider the case $\mu = 1/2$, which corresponds to the third regime in Eq. (\ref{eq:asympt_log}). In Fig. \ref{fig_precise_III} we show a plot of $p_{\rm logtime}(Y)\, \exp{(\gamma \, Y)}$, for $\mu = 1/2$ and $\gamma =1/2$ as a function of $Y$ and for different durations of the CTRW, $T = 10^4$, $T=4\,\cdot 10^4$, and $T = 10^5$. As before, the data have been obtained by averaging over $10^6$ independent CTRWs. The linear behavior, in a good agreement with our analytical prediction $\propto {\cal A}'_{\rm III} \, Y$, is a clear signature of the logarithmic correction of $p_{\rm time}(t)$ as expected from our analytical result in Eq. (\ref{eq2:result_p_of_t}).   

Finally, Fig. \ref{fig_precise_II} shows a plot of $p_{\rm logtime}(Y) \exp{(\gamma Y)}$ as a function of $Y$, for $\mu =1$, $\gamma = 1/2$, and the same walk durations as before (averaging procedure is also the same). It can be seen that our numerical data are compatible with the expected quadratic behavior\ (\ref{eq:asympt_log}), $p_{\rm logtime}(Y) \exp{(\gamma Y)} \sim {\cal A}'_{\rm II} \, Y^2$. Recalling that $Y = \log(t/\tau_c)$, this quadratic behavior confirms our analytical predictions for $p_{\rm time}(t)$ in Eq. (\ref{eq2:result_p_of_t}). 

%\newpage

%
%%%%%%%%%%%%%%%%%%%%
%
\section{Summary and perspectives}\label{sec9}
In this paper, we have performed a detailed analytical study of the statistics of the gap $g$ and the time interval $t$ between the first two maxima of a CTRW in the limit where the number of steps in the walk goes to infinity. The results we have obtained are quite general as we have addressed the question for the wide class of a symmetric, bounded, and (piecewise) continuous jump distribution, including the case of L\'evy flights of index $\mu$, with $0<\mu <2$. The average trapping time $\langle\tau\rangle$ between two successive jumps can be either finite, $\langle\tau\rangle =\tau_c <+\infty$, or infinite. In the latter case, the trapping time distribution $\Psi(\tau)$ is assumed to have a power law tail $\Psi(\tau) \propto \tau^{-1-\gamma}$, with $0 < \gamma <1$.

For a finite average trapping time between two successive jumps, our results coincide with the ones in\ \cite{MMS2014}, devoted to discrete time RWs, in which the number of jumps between the first two maxima, $l$, is merely replaced with $t/\tau_c$, as expected from simple law of large number arguments. On the other hand, our study provides a non trivial extension of the one in\ \cite{MMS2014} to the case where the average time between two successive jumps does not exist and law of large number arguments cannot be used. In particular, by taking $\gamma <\mu/2$ it makes it possible to address the question for sub-diffusive RWs, out of reach of the discrete time analysis of\ \cite{MMS2014}.

We have first shown that the joint PDF $p_n(g,t)$ converges to a stationary PDF, $p_n(g,t)\rightarrow p(g,t)$, as $n \to \infty$, where $n$ is the total number of steps in the walk. We have then obtained an explicit expression of the Laplace transform of $p(g,t)$ w.r.t. $t$, given in equations (\ref{eq2.8}) and (\ref{eq2.5}), from which we have performed a detailed analysis of $p(g,t)$ in the plane $(g,t)$ for $\mu$ and $\gamma$ in the whole ranges $0 < \mu \leq 2$ and $0<\gamma\le 1$, and for the three different main classes identified in\ \cite{MMS2014}: (i) slow, (ii) exponentially, and (iii) fast decreasing $f(\eta)$ at large $\eta$.

In the case (iii) of a fast decreasing jump distribution ($\mu =2$) with $0<\gamma <1$ and in the scaling regime $g,\ t \gg 1$ with fixed $t^{\gamma/2} g^{-\mu}$, we have shown that $p(g,t)$ takes a scaling form, Eqs.\ (\ref{scaling1a}) and\ (\ref{scaling1b}). The switch from the first to the second behavior\ (\ref{scaling1b}) corresponds to the cross-over from a `concentration' -- or `one-step' --  regime where the walker get stuck for a long time $t$ at the second maximum and then jumps directly to the first maximum, to a `many-steps' regime where she/he travels a long walk of total duration $t$ (with many steps) between the second and the first maxima. We have summarized these results in Figure\ \ref{fig1}. The numerical observation of the cross-over and `one-step' regime is still out of reach as it requires a prohibitively large value of $g$. Note, however, that we expect a similar scaling form for $p(g,t)$ if $f(\eta)$ has a bounded support $-\eta_{\rm max}\le\eta\le\eta_{\rm max}$ and $g\rightarrow +\infty$ is replaced with $g\rightarrow\eta_{\rm max}$, the largest possible (bounded) value of the gap. This should make a numerical observation of the cross-over and `one-step' regime easier in this case. The study of $p(g,t)$ for jump distributions with a bounded support (in both discrete and continuous time settings) will be the subject of a future work.

For type (i) L\'evy flights of index $\mu$, with $0 < \mu < 2$, and in the scaling regime $g,\ t \gg 1$ with fixed $t^\gamma g^{-\mu}$, we have shown that $p(g,t)$ takes the scaling form\ (\ref{scaling2a}) with the asymptotic behaviors\ (\ref{scaling2b}) and\ (\ref{scaling2c}). These results, summarized in Figures\ \ref{fig2a} to\ \ref{fig2c}, provide a non trivial extension of the ones in\ \cite{MMS2014} to the case $0<\gamma <1$.

Finally, from the stationary joint PDF $p(g,t)$ we have computed the stationary marginal distribution, $p_{\rm time}(t)$, of the time between the first to maxima of the walk. Its asymptotic behavior, which depends on $\gamma$ and the L\'evy index $\mu$, is summarized in equations\ (\ref{eq1:result_p_of_t}),\ (\ref{eq2:result_p_of_t}), and in Table\ \ref{tab1}. The third lines of\ (\ref{eq1:result_p_of_t}) and\ (\ref{eq2:result_p_of_t}) reveal that the freezing phenomenon of the large $t$ behavior of $p_{\rm time}(t)$, as a function of $\mu$, as $\mu$ decreases past the value $\mu_c = 1$, which was first put in evidence for discrete time RWs in\ \cite{MMS2013,MMS2014}, also exists for CTRWs. As a consequence, the first moment of $p_{\rm time}(t)$ is never defined.

Our study raises interesting open questions. For instance, this work focuses on the first gap and it would be interesting to study higher order gaps. In
Ref. \cite{SM2012}, the study of the $k$-th gap of long random walks, for $\mu =2$ revealed a rich and universal behavior (i.e., independent of the jump distribution) of the stationary PDF of the $k$-th gap in the large $k$ limit. The extension of these results, which also hold for 
CTRW with $\mu =2$ and any value of $0<\gamma \leq 1$, to L\'evy flights, i.e. $0 < \mu < 2$, remains a challenging open question. 
Similarly, it would very interesting to study the time between the $k$-th and $(k+1)$-th maximum for CTRW (both for $\gamma = 1$ and $\gamma <1$), for which one also expects universality for $k \gg 1$. Finally, the generalization of our results to higher dimensions would also be very interesting.

%%%%%%%%%%%%%%%%%%%%%%%%%%%%%%%%%%%%%
%

%
%
\end{document}